\documentclass[preprints,article,accept,pdftex,moreauthors]{Definitions/mdpi} 
\RequirePackage{siunitx}
\RequirePackage{subcaption}
\usepackage[acronym, section=section]{glossaries}
\makeglossaries

\newglossaryentry{Qbunch}{
  name={\ensuremath{Q_\text{b}}},
  description={Bunch charge}
}
\newglossaryentry{sigmab}{
  name={\ensuremath{\sigma_\text{b}}},
  description={Longitudinal width $\sigma$ of the bunch}
}
\newglossaryentry{sigmabtrans}{
  name={\ensuremath{\sigma_\text{b,trans}}},
  description={Transversal width $\sigma$ of the bunch}
}
\newglossaryentry{beta}{
  name={\ensuremath{\beta}},
  description={Percentage of the speed of light}
}
\newglossaryentry{c0}{
  name={\ensuremath{c_0}},
  description={Speed of light}
}
\newglossaryentry{vb}{
  name={\ensuremath{v_\text{b}}},
  description={Velocity of the beam}
}
\newglossaryentry{ar10+}{
  name={\ensuremath{\text{Ar}^{10+}}},
  description={Argon ion with 10 stripped electrons}
}
\newglossaryentry{Ne}{
  name={\ensuremath{N_\text{e}}},
  description={Number of secondary electrons}
}
\newglossaryentry{Nion}{
  name={\ensuremath{N_\text{ion}}},
  description={Number of secondary electrons}
}
\newglossaryentry{ne}{
  name={\ensuremath{n_\text{e}}},
  description={Electron density of the collector material}
}
\newglossaryentry{z}{
  name={\ensuremath{Z}},
  description={Ion charge state}
}
\newglossaryentry{sigmac}{
    name={\ensuremath{\sigma_\text{c}}},
    description={Calculated bunch length including the effect of the gap between chassis and collector of a radially coupled FFC}
}
\newglossaryentry{sigmas}{
    name={\ensuremath{\sigma_\text{s}}},
    description={Evaluation of the bunch length using the base length of the signal}
}
\newglossaryentry{sigmagap}{
    name={\ensuremath{\sigma_\text{gap}}},
    description={Expansion of \gls{sigmab} due to the distance between chassis and collector}
}
\newglossaryentry{tsep}{
    name={\ensuremath{t_\text{sep}}},
    description={temporal separation time between incoming ion signal from the secondary electron signal}
}

\newacronym{ffc}{FFC}{Fast Faraday cup}
\newacronym{rcffc}{RCFFC}{radially coupled Fast Faraday cup}
\newacronym{trcffc}{TRCFFC}{tapered radially coupled Fast Faraday cup}
\newacronym{acffc}{ACFFC}{axially coupled Fast Faraday cup}
\newacronym{se}{SE}{secondary electron}
\newacronym{see}{SEE}{secondary electron emission}
\newacronym{sey}{SEY}{secondary electron yield}
\newacronym{cst}{CST}{CST Studio Suite\texttrademark}
\newacronym{T}{T}{turned}
\newacronym{cm}{CM}{conventionally machined}
\newacronym{AM}{AM}{additively manufactured}
\newacronym{lpbf}{LPBF}{laser powder bed fusion}
\newacronym{iacs}{IACS}{international annealed copper standard}
\newacronym{gsi}{GSI}{GSI Helmholtz Centre for Heavy Ion Research}

\firstpage{1} 
\pubvolume{1}
\issuenum{1}
\articlenumber{0}
\pubyear{2025}
\copyrightyear{2025}
\datereceived{ 6 Aug 2025 } 
\daterevised{ } 
\dateaccepted{ } 
\datepublished{ } 
\hreflink{MDPI Manuscript ID: instruments-3833241} 

\Title{\uppercase{RF characterization and beam measurements with 3D printed Fast Faraday cups}}

\TitleCitation{\uppercase{RF characterization and beam measurements with 3D printed Fast Faraday cups}}

\Author{Stephan Klaproth $^{1,4,\ddagger}$\orcidA{}*,
        Rahul Singh $^{2,\ddagger}$\orcidB{},
        Samira Gruber$^{3}$\orcidC{},
        Lukas Stepien$^{3}$\orcidD{},
        Herbert De Gersem$^{4}$\orcidE{} and
        Andreas Penirschke $^{1}$\orcidF{}}

\address{%
$^{1}$ \quad Technische Hochschule Mittelhessen, Friedberg, Germany; Stephan.Klaproth@iem.thm.de\\$^{2}$ \quad GSI Helmholtzzentrum für Schwerionenforschung, Darmstadt, Germany; R.Singh@gsi.de\\
$^{3}$ \quad Fraunhofer Institut für Werkstoff- und Strahltechnik IWS, Dresden, Germany; samira.gruber@iws.fraunhofer.de, lukas.stepien@iws.fraunhofer.de\\
$^{4}$ \quad Technische Universität Darmstadt, Darmstadt, Germany; degersem@temf.tu-darmstadt.de}

\corres{Correspondence: Stephan.Klaproth@iem.thm.de; Tel.: +49 641/309-3223 (S.K.)}

\secondnote{These authors contributed equally to this work.}

\abstract{ 
The early stages of most particle accelerator chains produce sub-ns bunches with velocities in the range of \SIrange{1}{20}{\percent} of the speed of light. \glspl{ffc} are designed to measure the longitudinal charge distribution of these short bunches of free charges. Coaxial designs have been utilized at the \gls{gsi}'s linear accelerator UNILAC  \cite{Barth2022} to characterize ion bunches with bunch lengths ranging from  a few hundred ps to a few ns. The typical design goals are to avoid the pre-field of the charges and to suppress \gls{see}, while retaining the capability of bunch-by-bunch measurements. In this contribution, a novel \gls{ffc} design  manufactured using additive manufacturing, e.g. \gls{lpbf} is presented and compared with a traditionally produced \gls{ffc}. The design considerations, RF characterization, and selected measurements with ion beam at \gls{gsi} are shown.
}


\begin{document}

\section{Introduction}
\glsresetall
In accelerator instrumentation, Faraday cups represent a class of diagnostic devices specifically designed to directly collect and quantitatively measure the charge deposited by incident particle beams. Faraday cups provide absolute measurements of beam current and are widely known for their robustness, simplicity and reliability in a variety of accelerator applications.

Faraday cups modified to measure the fast temporal evolution of charges are referred to as \glspl{ffc}. The temporal evolution of the measured bunch charge \gls{Qbunch} or longitudinal charge distribution is also referred to as the shape of the bunch. In literature, many different \gls{ffc} designs have been proposed, such as stripline, microchannel, multi-aperture, planar, and conical geometries \cite{Strehl2006,Bogaty1990, Rawnsley2001,Carneiro2019, Ferianis2003,Mathew2020,Mal2022}. Regardless of the RF signal guiding mechanism or form factors, all \glspl{ffc} are designed as broadband detectors to achieve the necessary temporal resolution for their specific use case. For bunches at the linear accelerator UNILAC of \gls{gsi}, bunch lengths \gls{sigmab} of a few hundred \si{\pico\second} to a \si{\nano\second} are observed, and accordingly, a bandwidth of \SIrange{1}{5}{\giga\hertz} is aimed for. This bandwidth regime can be conveniently covered by coaxial \gls{ffc} designs. The primary challenge in measuring the charge distribution of particles with beam velocities $\gls{beta} < \SI{100}{\percent}$ lies in the significant alteration of both, the charge distribution and its electromagnetic field distribution caused by Lorentz boost and transformation \cite{Jackson1999}. This effect becomes especially prominent for slow beams ($\gls{beta} < 0.25$) with short bunches ($ < ns$).

A common method for mitigating the influence of the pre-field on the detector signal is to add electrostatic shielding such as fine-mesh grids \cite{Rawnsley2001} or small-aperture pinholes \cite{Carneiro2019,Mal2022}. The second challenge for \glspl{ffc} is the effective suppression of \glspl{se}, which are emitted both by the direct impact of impinging ions and by the interaction with high-energy \glspl{se} generated within the detector materials. The presence of \glspl{se} can significantly alter the measured bunch shape, making robust suppression mechanisms indispensable for retrieving the true, undisturbed bunch shape. Established methods include field-based suppression as well as geometric mitigation: field-based suppression involves the application of electric \cite{Rawnsley2001} or magnetic fields to confine the \gls{se} within the \gls{ffc}. Geometric mitigation modifies the geometry of the collector of the \gls{ffc}, e.g., conically, to redirect the main direction of the emitted \gls{se} toward surfaces where they are rapidly reabsorbed \cite{Masoumzadeh2019}.

This contribution begins with a systematic analysis of axially and radially coupled coaxial \glspl{ffc} with respect to their RF-characteristics and secondary electron suppression performance as simulated in \gls{cst} \cite{CSTStudioSuite}. Subsequently, an optimized tapered coaxial \gls{trcffc} design tailored to our use cases, is presented. Detailed RF simulations are shown along with insights on the manufacturing and construction. This design is found to have a significant improvement in secondary electron suppression and signal-to-noise ratio for bunch-by-bunch measurements at the cost of a reduction in bandwidth. Finally, some beam measurements carried out at the \gls{gsi} experimental cave X2 comparing the traditional and new designs are presented.

\subsection{Working principle of \gls{acffc}}

\begin{figure}[!htb]
    \centering
    \begin{subfigure}{0.56\textwidth}
        \includegraphics*[width=.9\columnwidth]{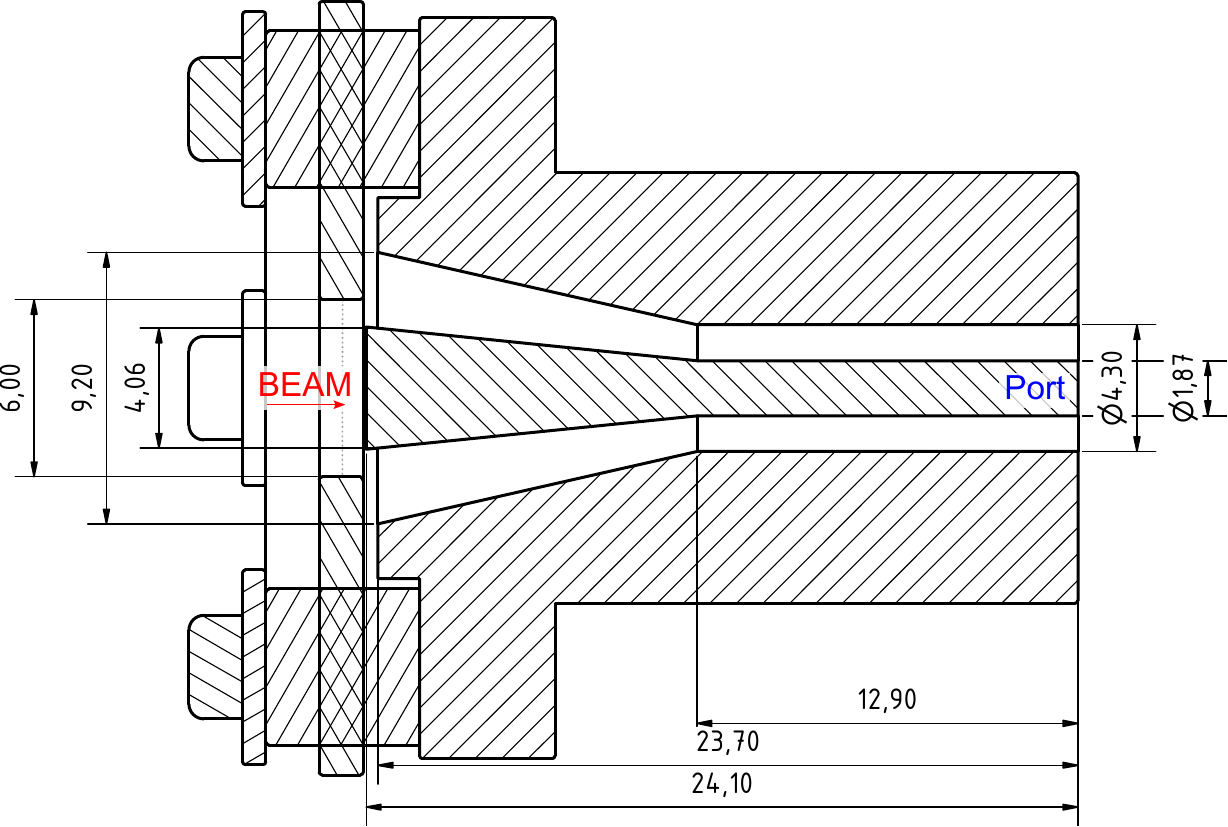}
        \caption{}
        \label{fig-001a}
    \end{subfigure}
    \hfill
    \begin{subfigure}{0.39\textwidth}
        \includegraphics*[width=\columnwidth]{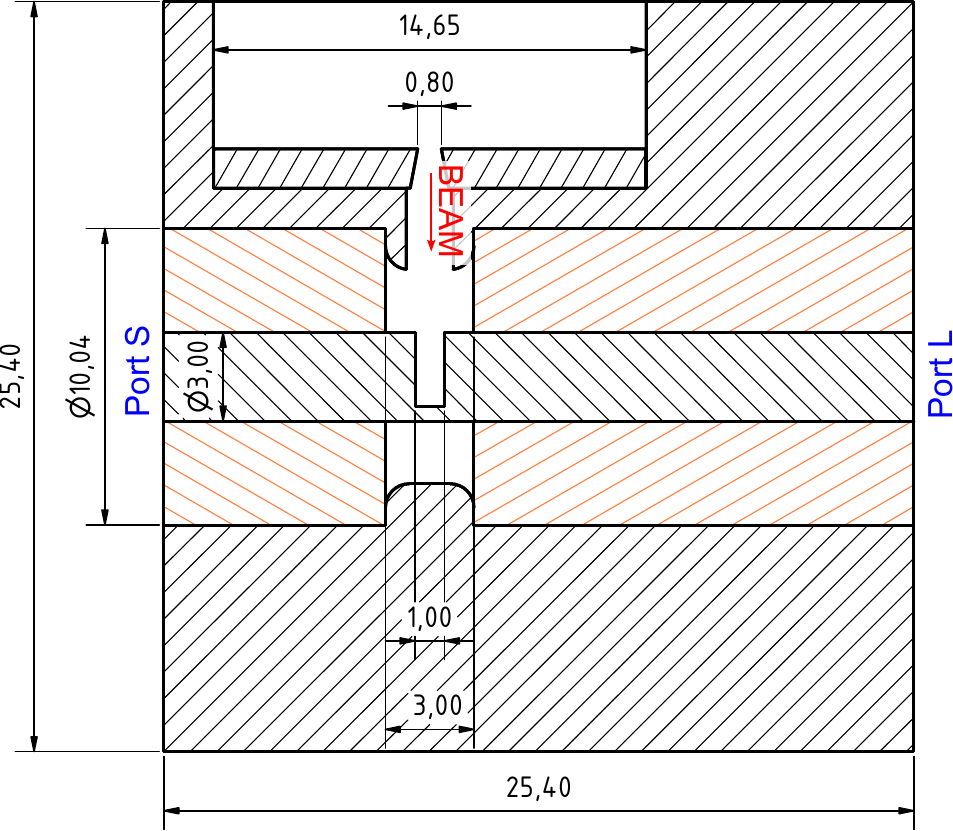}
        \caption{}
        \label{fig-001b}
    \end{subfigure}
    \caption{Cross section view of an (\subref{fig-001a}) axially coupled \gls{ffc} \cite{Rawnsley2001} and (\subref{fig-001b}) a radially coupled \gls{ffc} \cite{Carneiro2019}.}
    \label{fig-001}
\end{figure}

\begin{figure}[!htb]
    \centering
    \includegraphics*[width=.95\columnwidth]{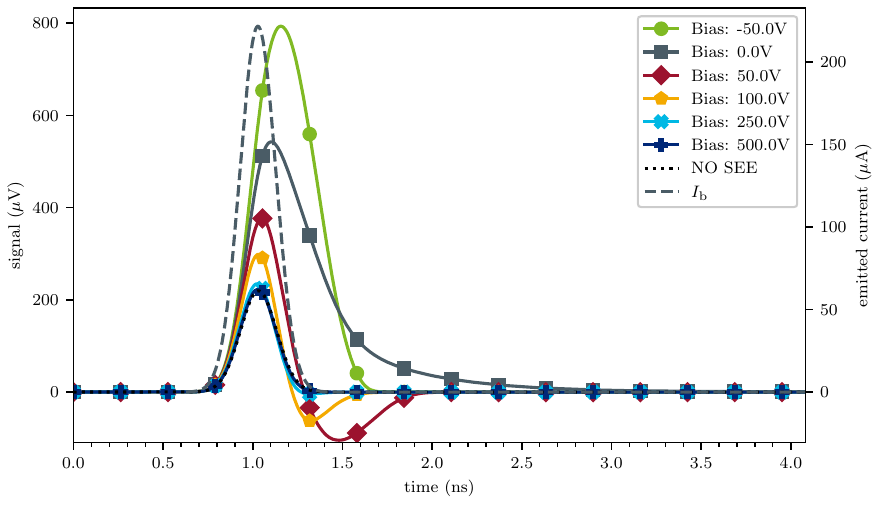}
    \caption{\gls{cst} PIC simulation of the \gls{acffc} (\ref{fig-001a}) with $\gls{sigmab}=\SI{100}{\pico\second}$ and different bias settings with activated \gls{see}.}
    \label{fig-002}
\end{figure}

Axially coupled coaxial FFCs were first documented in literature in the 1980s \cite{Fujimoto1982,Chuaqui1989}, employing coaxial geometries where the beam deposits charge on the inner conductor of the tapered end of the coaxial line. A representative commercial design \cite{Rawnsley2001} of this type of \gls{ffc} is shown in Fig.~\ref{fig-001a}, optimized to ensure that the beam interaction occurs exclusively with the collector, being the inner conductor of the \gls{ffc}. Usually, this is achieved with a collimator in front of the grid to limit the transversal beam width. The collimator is removed in the simulations to reduce the complexity of the model, though the transversal beam width can be controlled in the simulation. The grid in front of the collector, with a distance of \SI{0.85}{\milli\metre}, is used to apply an electrical field between the collector and the grid to suppress the secondaries and shield the pre-field of the bunch. A single Gaussian bunch of \num{34680} \gls{ar10+} ions was simulated with a bunch length of $\gls{sigmab}=\SI{100}{\pico\second}$ and a velocity $\gls{vb}=0.15\cdot \gls{c0}$. The collector is set to a static potential, while the grid is kept at ground. In this configuration, the collector attracts the \gls{se}. The effect of the \gls{se} can be studied in the simulations by varying the potential, as shown in Fig.~\ref{fig-002}. This contribution focuses only on simulation results and refers to \cite{Singh2022} for measurement results using this type of \gls{ffc}.

In the simulations, a solely ion-induced \gls{se} emission originating from the collector is allowed. The secondary electron yield and corresponding energy spectra have been extensively investigated in the literature \cite{Sternglass1957,Rothard1990,Koyama1981,Hasselkamp1981}, with recent research focusing on their application in profile grids for transverse ion beam profile measurements \cite{Reiter2012}. The approximated backscattered \gls{se} yield per ion can be derived from the Bethe-Bloch formula, which quantifies the stopping power of a charged particle beam interacting with a target. Consequently, the number of \glspl{se} \gls{Ne} scales with the square of the incident ion's charge state \gls{z} and the target electron density \gls{ne} of the \gls{ffc} material \cite{Singh2022}:
\begin{linenomath}
    \begin{equation}
        \gls{Ne} \propto \gls{ne} \gls{z}^2 \cdot \gls{Nion} \,.
    \end{equation}
    \label{eq-3}
\end{linenomath}
To achieve maximum \gls{sey}, the energy dependency of the Bethe-Bloch formula is disregarded, although the kinetic energy of the ions exceeds the threshold energy $E_\text{threshold}$ to emit \gls{se} of copper \cite{Haque2019,Benka1995}.
In this worst-case scenario, a \gls{sey} of $\gls{Ne}=\num{50}$ \gls{se} per ion is estimated for an \gls{ar10+} beam.
The value matches with empirical estimates reported in the literature of about 10-15 electrons per \SI{8.0}{\mega\electronvolt\per u} $O^{5+}$ ion incident to the normal incident on a copper target \cite{Koyama1981,Benka1995,Haque2019}.
In the imported emission model of \gls{cst}, the \gls{sey} curve for the normal incidence ($\Theta=0$) is defined by the user and will be estimated to  \gls{Ne}. The angular dependence of \gls{se} emission is internally calculated by \gls{cst} using the relations given in Eq.~(\ref{eq-6}, \ref{eq-7}) retrieving the following equations for the angular energy $E(\Theta)$ and the number $\delta(\Theta)$ of the \gls{se} \cite{CST2024, CSTStudioSuite}:
\begin{linenomath}
    \begin{equation}
        E(\Theta) = E_0 - E_\text{threshold} \frac{2\pi}{2\pi + \Theta^2} + E_\text{threshold} \,;
        \label{eq-6}
    \end{equation}
    \begin{equation}
        \delta(\Theta) = \delta(0) \left(1+\frac{\Theta^2}{2\pi}\right) \,,
        \label{eq-7}
    \end{equation}
\end{linenomath}
where $\delta(0)$ is the \gls{sey} at normal incidence ($\Theta=0$) and $E_0$ is the incident ion energy.
Surface roughness is neglected by \gls{cst} in the imported emission model \cite{CST2024, CSTStudioSuite}, resulting in the \gls{sey} angular emission being treated as proportional to $\sec(\Theta)$, as it is characteristic for smooth surfaces \cite{Hasselkamp1992}:
\begin{linenomath}
    \begin{equation}
        f(E) = \delta\left(E_0,\Theta_0\right) \frac{E}{T_e^2} e^{-\frac{E}{T_e}} P^{-1}\left(2,\frac{E_0}{T_e}\right) \,,
        \label{eq-8}
    \end{equation}
\end{linenomath}
with the incomplete gamma distribution $P$, the most common energy $T_e$ (also denoted as temperature of the \gls{se}) and the energy of the secondary electron $E$. The energy distribution $f(E)$ of the \gls{se} for $E_\text{threshold}=0$ is equal to the Vaughan model \cite{Vaughan1993}. In this case, the \gls{sey} $\delta\left(E_0,\Theta_0\right)$ becomes dependent on the incident angle $\Theta_0$ and the incident energy $E_0$ through the relations Eq.~(\ref{eq-6}, \ref{eq-7}).
Typically, the temperature of \gls{se} is in the range of \SIrange{5}{10}{\electronvolt} \cite{Sternglass1957,Rothard1990}. To represent a worst-case scenario regarding the influence of the \gls{se}, the $T_e$ was set to \SI{10}{\electronvolt} in the simulations. Consequently, for an electron temperature $T_e$ of \SI{10}{\electronvolt}, \SI{96}{\percent} of the \gls{se} have energies below \SI{50}{\electronvolt} .
Depending on the bias voltage, this percentage represents the number of \gls{se} that return to the collector. It should be noted that the secondaries are emitted from the conductor, propagate into free space, and, after being decelerated, subsequently regain velocity toward the collector. The process introduces a temporal delay due to the travel time of the \gls{se}. The extent to which the bias voltage exceeds the energy of the \gls{se} determines how quickly the point of return is reached.
The simulation results of the design depicted in Fig.~\ref{fig-001a} are presented in Figure~\ref{fig-002}, with bias voltages configured between \SI{-50}{\volt} and \SI{100}{\volt}. The emitted current is shown as a dashed line alongside the output signals, while the output obtained with deactivated \gls{se} emission included as a dotted line, serving as a reference baseline. For all simulation steps, the pre-field of the bunch is shielded by a grid, minimizing the elongation caused by it to a few \si{\pico\second} before the actual charges reach the collecting surface.
A total of three different regimes can be identified within the bias schemes.

In the first regime, the collector is negatively biased. As a result, the \glspl{se} are pushed away from the collector and do not return to it. As the \gls{se} move further away from the collecting surface, their electric field terminates only partially on the collector, with an  increasing portion of the field terminating on the shielding grid or chassis. The mirror charges, which are no longer bound by the electric field of the \gls{se} on the collector, are able to move and thus create an additional positive signal. The propagation velocity of the \gls{se} toward the grid is directly proportional to their kinetic energy. Consequently, a larger portion of their field is terminated on the grid more rapidly. This results in an additional positive signal from the mirror charges which is superimposed onto the ion-induced signal with minimum delay, ultimately increasing both signal amplitude and signal width. Furthermore, the influence of \gls{se} on the measured signal is strongly dependent on the speed of the \gls{se} and, thus, on the temperature $T_e$. While the actual energy distribution of the \gls{se} is of no significance for high negative bias settings w.r.t. the electron temperature, it becomes more obvious looking at (\SI{0}{\volt}) under unbiased conditions. The clearly visible long tail reflects the energy distribution of the \gls{se}. The fast \gls{se} mainly increase the total signal strength, while the slow \gls{se} primarily extend the tail.

The second regime is characterized by the application of a moderate positive bias. In this regime, the \gls{se} have enough energy to travel a considerable distance from the collecting surface, allowing the mirror charges in the collector to be partially transferred to the output. This initially results in an increase in the signal strength. Then, \gls{se} with less energy than the bias voltage return, creating a negative output signal without generating a significant number of tertiary electrons due to their low energy. These returning \gls{se} are partially compensated by the mirror charges of newly generated \gls{se} until the production of new \gls{se} equals the returning \gls{se}. Any additional \gls{se} beyond the compensated ones contribute to a negative time-delayed signal, ultimately causing the underswing after the main peak.

The third regime refers to high biasing ($>$\SI{200}{\volt}). In this case, the \gls{se} do not travel far into free space. Consequently, the mirror charges remain stationary and recombine with the \gls{se}, resulting in no change in either signal strength or signal width. Typically, a small portion of high-energy \gls{se} still reach far enough so that a tiny underswing remains.

\subsection{Working principle of \gls{rcffc}}
\begin{figure}[!htb]
    \centering
    \begin{subfigure}{0.49\textwidth}
        \includegraphics*[width=.95\columnwidth]{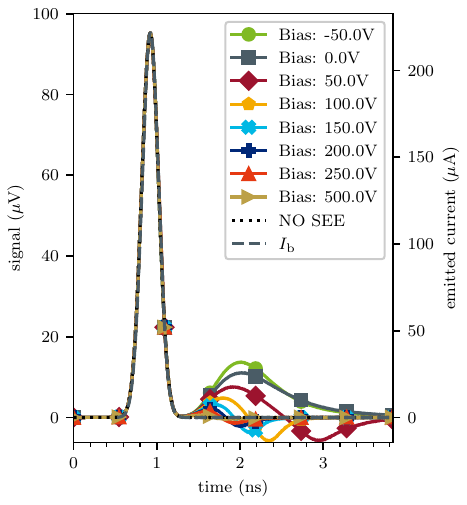}
        \caption{}
        \label{fig-003a}
    \end{subfigure}
    \hfill
    \begin{subfigure}{0.49\textwidth}
        \includegraphics*[width=.95\columnwidth]{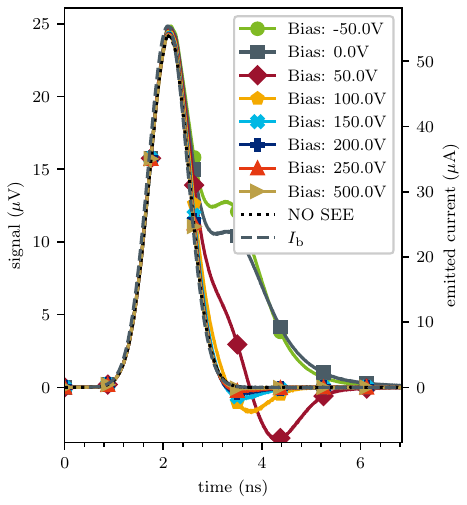}
        \caption{}
        \label{fig-003b}
    \end{subfigure}
    \caption{\gls{cst} PIC simulations of the \gls{rcffc} (\ref{fig-001b}) with different bias settings and bunch lengths (\subref{fig-003a}) $\gls{sigmab}=\SI{100}{\pico\second}$ and (\subref{fig-003b}) $\gls{sigmab}=\SI{400}{\pico\second}$ with activated \gls{see}.}
    \label{fig-003}
\end{figure}

Effective suppression of \gls{se} based solely on a bias scheme is challenging.
In general, it is advisable to combine different methods to suppress \gls{se} reducing the impact on the output signal as much as possible.
An example of combined biasing and geometrical suppression of \gls{se} is provided by the second design shown in Fig.~\ref{fig-001b}, which features a radially coupled \gls{ffc}.
The collector of the \gls{rcffc} is fabricated with an outer diameter of \SI{3}{\milli\metre} and contains a radially drilled hole with a diameter of \SI{2}{\milli\meter} and a depth of \SI{2.5}{\milli\metre}. In the simulation, the bottom of the drill hole is modeled as flat, whereas in practice, drills typically have a point angle of \SIrange{115}{138}{\degree}, resulting in a short, conical end of the hole. The pre-field of the bunch is shielded by a $\varnothing \SI{0.8}{\milli\meter}$ pinhole in the heat shield, so that for the simulation, only the distance between the collector and the outer conductor of maximum \SI{3}{\milli\meter} is relevant. Around the drill hole, there is a reduced diameter section of the outer conductor to further reduce the critical distance to \SI{2}{\milli\meter}. This also helps to generate higher field strengths, so that the \gls{se} will return earlier and match the geometry to \SI{50}{\ohm}. In Figure~\ref{fig-003}, the bias scheme for the \gls{rcffc} for two different bunch lengths is shown.

An \gls{rcffc} has two ports instead of one for the \gls{acffc}, splitting the signal equally. Hence, the maximum amplitude of the ion signal of the \gls{acffc} (see Fig.~\ref{fig-002} case "NO SEE" and Bias \SI{500}{\volt}) per port is halved. On the other hand, two ports allow for different noise suppression schemes during the signal processing.
For short bunches (Fig.~\ref{fig-003a}), the ions are detected without any deviation from the baseline without SEE matching the exact ion current regardless of the bias used. Only the signal part of the \gls{se} after the ion peak is bias-dependent. The number of detected \gls{se} is reduced by a factor of about \num{25} compared to the \gls{acffc}. Both, the separation of the \gls{se} from the ion signal and the reduction of the \gls{se} are the result of the geometrical \gls{se} suppression of the drill hole in the collector.
The electric fields of the \gls{se} terminate on the drill hole walls as long as the \gls{se} remains inside the drill hole.
The mirror charges move upward along with the \gls{se} and remain bound to them until the \gls{se} leave the drill hole.

\begin{figure}[!htb]
    \centering
    \includegraphics*[width=.95\columnwidth]{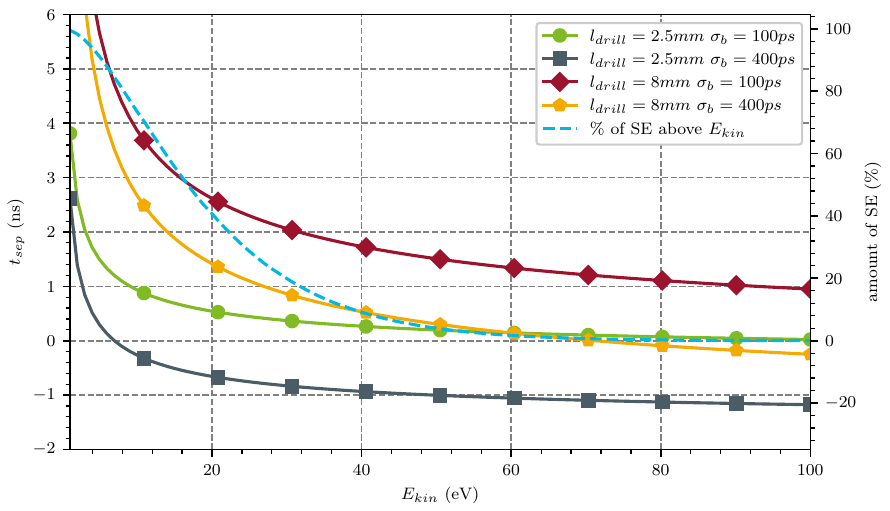}
    \caption{Temporal separation of the \gls{se} peak from the ion peak for different long drill holes and bunch lengths. In addition, the fraction of \gls{se} of the gamma distribution for $T_e=\SI{10}{\electronvolt}$ above the respectively kinetic energy $E_\text{kin}$ is shown as dashed line.}
    \label{fig-004}
\end{figure}

If the bunches become longer (Fig.~\ref{fig-003b}), both, the signal of the ions and the \gls{se} overlap. However, the \gls{se} peak is significantly delayed after the ion peak.
The remaining increase of the ion peak due to the \gls{se} is still negligible in this case. The effect of the \gls{se} can still be suppressed taking the biasing into account.
For even longer bunches, the benefit of the separation is lost and only the reduction of the total number of \gls{se} remains. The pinhole in the heat shield strongly reduces the signal amplitude compared to an \gls{acffc}, while relying primarily on the bias scheme for suppression. Therefore, the drill hole depth should be adjusted to the expected bunch length.
According to the gamma-distributed \gls{se}, the velocity $v_e$ of \SI{96}{\percent} ($<\SI{50}{\electronvolt}$) of \gls{se} is lower than \SI{4.2}{\milli\metre\per\nano\second}
. The separation time $t_\text{sep}$ for each velocity $v_e$ can be calculated for a given expected bunch length \gls{sigmab} to optimize the drill hole depth of an \gls{rcffc}:
\begin{linenomath}
    \begin{equation}
        \gls{tsep}= \frac{l_\text{drill}}{v_e}-4\gls{sigmab} \,.
    \end{equation}
\end{linenomath}
Figure~\ref{fig-004} shows the separation time $t_\text{sep}$ for different kinetic energies and bunch lengths.
The dashed line represents the cumulative portion of \gls{se} with a kinetic energy above $E_\text{kin}$.
If the separation time is positive, \gls{se} with energy specified or lower are separated from the ion signal. Conversely, if the separation time is negative, all \gls{se} with kinetic energies above this threshold will overlap with the ion peak.
The \gls{rcffc} shown in Fig.~\ref{fig-001b} has a drill hole depth of \SI{2.5}{\milli\metre}. For a bunch length of $\gls{sigmab}=\SI{100}{\pico\second}$, no overlapping \gls{se} are expected, which aligns with the simulation results in Fig.~\ref{fig-003a}. However, for a bunch length of $\gls{sigmab}=\SI{400}{\pico\second}$, \gls{se} with an energy above \SI{6.9}{\electronvolt} (\SI{85}{\percent} of all \gls{se}) will leave the drill hole too early, causing the signal of the \gls{se} to overlap, as shown in Fig.~\ref{fig-003b}. Increasing the drill hole depth to \SI{8}{\milli\metre} ensures that the signal of \gls{se} with a kinetic energy below \SI{71.1}{\electronvolt} (\SI{99.36}{\percent} of all \gls{se}) remains separated, with only rare fraction (\SI{0.64}{\percent}) of faster \gls{se} overlapping. Thus, adjusting the drill hole depth depending on the expected bunch lengths helps to reduce the impact of the \gls{se} on the measured signal.

\section{3D printed Tapered Radially coupled \gls{ffc}}
In this section, details regarding the layout and design choices for the \gls{trcffc} are provided, and the differences between conventionally machined and additively manufactured collectors are discussed. The comparisons are supported by  simulations and measurements of the \gls{trcffc}'s RF-properties. The final design is compared to a \gls{rcffc} measuring an \gls{ar10+} particle beam at the \gls{gsi} linear accelerator UNILAC measured in the beam line X2.

\subsection{Tapered Radially coupled \gls{ffc} design considerations and simulations}
\label{Tapered_Radially_coupled_ffc_design_considerations_and_simulations}
\begin{figure}[!htb]
    \centering
    \includegraphics*[width=.8\columnwidth]{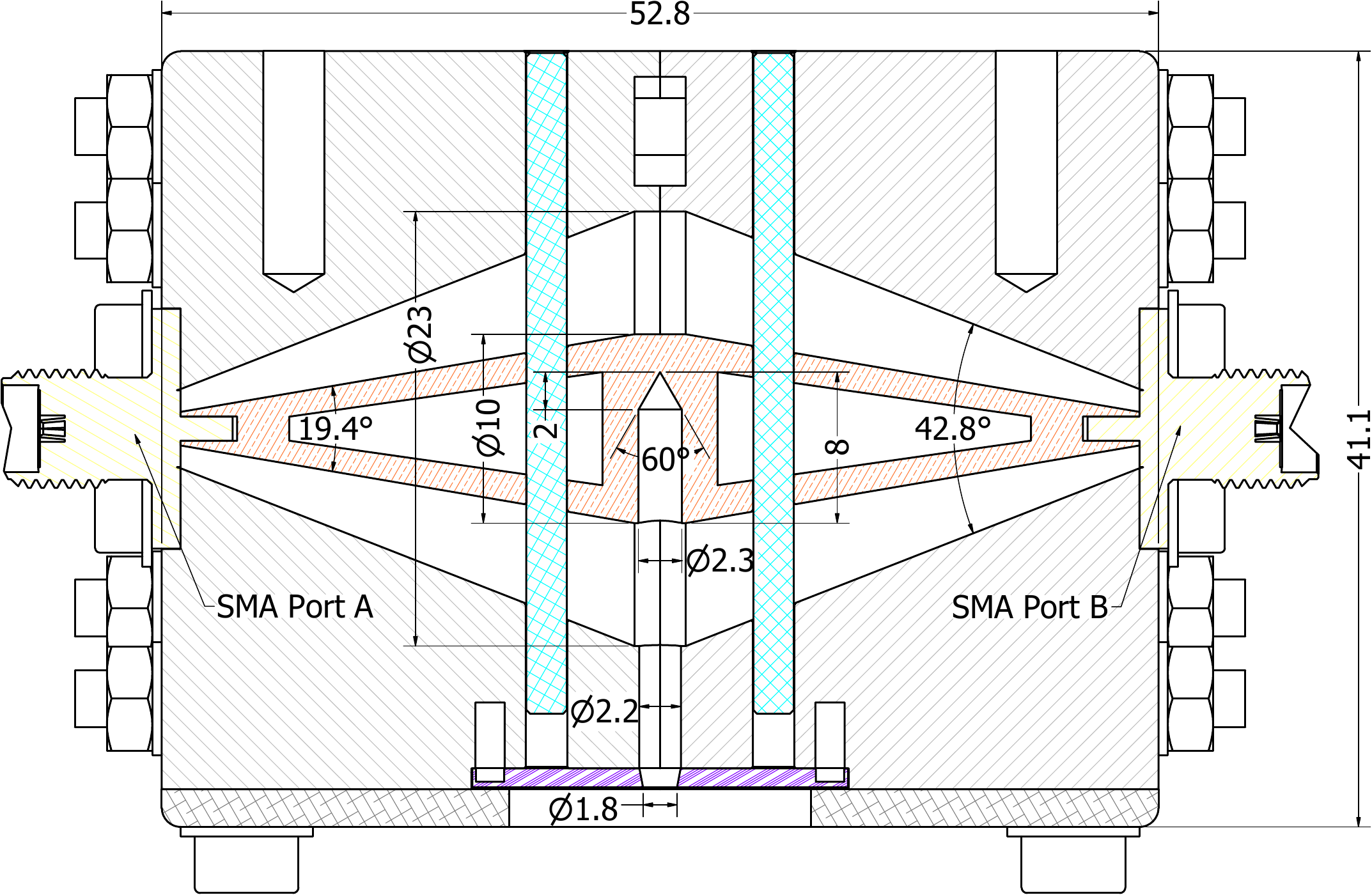}
    \caption{Cross section of a \gls{trcffc} with optimized geometrical \gls{se} suppression.}
    \label{fig-005}
\end{figure}

A new variant of the previously discussed \gls{rcffc} was developed for UNILAC at \gls{gsi}. The optimization objectives are defined as achieving a higher signal-to-noise ratio and enhancing the geometric suppression of \gls{se} for bunch lengths between \SIrange{100}{400}{\pico\second} at velocities ranging from \SIrange{5}{15}{\percent} of the speed of light. The new design is shown in Fig.~\ref{fig-005}. The diameter of the pinhole in the heat shield has been increased from $d_\text{RCFFC}=\SI{0.8}{\milli\metre}$ to $d_\text{TRCFFC}=\SI{1.8}{\milli\metre}$.
Comparing different pinhole sizes, a theoretical signal gain of \num{5.1} is achievable for a uniform distribution. This gain is determined by the ratio of the pinhole areas in the two designs.
The projected gain is reduced when considering the Gaussian profile of the transversal bunch shape, as discussed in Eq.~(\ref{eq-2}). Assuming a transversal beam size of $\gls{sigmabtrans}=\SI{2}{\milli\metre}$, the actual gain is approximately a factor of \num{2.0}.
\begin{linenomath}
    \begin{equation}
        f_\text{gain}  = \frac{\text{erf}\left(\frac{1}{\sqrt{2}}\frac{d_\text{TRCFFC}}{\gls{sigmabtrans}}\right)}{\text{erf}\left(\frac{1}{\sqrt{2}}\frac{d_\text{RCFFC}}{\gls{sigmabtrans}}\right)}
        \label{eq-2}
    \end{equation}
\end{linenomath}
The drift passage width after the pinhole in the heat shield is larger than the pinhole itself, ensuring that entering particles reach the drill hole without hitting the walls. Consequently, the drill hole in the collector needs to have a larger diameter (\SI{2.3}{\milli\metre}) than the drift passage to prevent generation of secondaries at the surface of the collector aside inside the drill hole.

\begin{figure}[!htb]
    \centering
    \includegraphics*[width=.75\columnwidth]{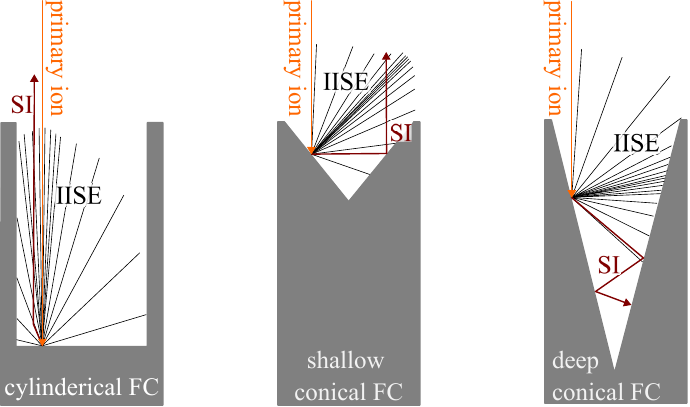}
    \caption{A simple model for ion-induced \glspl{se} (IISEs) and scattered ion (SI) emission due to incidence of a primary ion inside the Faraday cup for three different cavities \cite{Masoumzadeh2019}.}
    \label{fig-006}
\end{figure}

The number of \gls{se} leaving can be controlled by the depth-to-aperture ratio of the drill hole \cite{Mal2022}. This ratio is \num{2.5} for the \gls{rcffc} design. A minimum drill hole depth of \SI{5.75}{\milli\metre} is needed to reach at least the same depth-to-aperture ratio for a drill hole diameter of \SI{2.3}{\milli\metre}.
However, at these values, \SI{11.75}{\percent} of the leaving \gls{se} will not be separated from the ion-induced signal for $\gls{sigmab}=\SI{400}{\pico\second}$. Thus, a drill hole depth of \SI{8}{\milli\meter} was chosen, with the last \SI{2}{\milli\meter} featuring a \SI{60}{\degree} cone-shaped end.
This cone-shaped termination of the drill hole facilitates the redirection of \gls{se} emission toward the hole walls, as described by Eq.~(\ref{eq-7}) and (\ref{eq-8}). In Figure~\ref{fig-006}, a study of various Faraday cup geometries is presented, enabling to anticipate the effect of different drill hole terminations such as flat or wide and narrow cone-shaped on the \gls{se} suppression to be assessed.
While a flat-ended drill hole results in a main direction of the \gls{se} backward out of the hole, the cone-shaped tips will considerably increase \gls{se} absorption at the hole walls.

A more detailed analysis with respect to the angle of the cone shape has been performed in \cite{Mal2022}. It was found that a flat-tip design allows approximately \SI{20}{\percent} of the total \gls{se} to escape, whereas a cone-shaped tip with a \SI{56}{\degree} angle reduces this to a minimum of \SI{3.3}{\percent}. Applied to the present design featuring a  \SI{60}{\degree} cone, it is expected that approximately \SI{2.4}{\percent} of all emitted primary \gls{se} will escape the drilled hole. Of these, \SI{96}{\percent} are anticipated to be temporally separated from the ion peak due to their travel time through the drill hole.

The penetration depth $s_\text{pen}$ of the ions has been estimated using SRIM \cite{SRIM2013} for typical energies at the \gls{gsi} linear accelerator UNILAC ranging from \SIrange{1.4}{11.4}{\mega\electronvolt\per u} \cite{Barth2022} for a wide variety of ion species from H$_2$ to $^{238}$U in copper.
The results indicate that for uranium the penetration depth is $s_{\text{pen,}^{238}\text{U,11.4MeV/u}}=\SI{43.11}{\micro\metre}$ and for hydrogen $s_{\text{pen,H}_2\text{,11.4MeV/u}}=\SI{45.9}{\micro\metre}$. Accordingly, it is expected that any ion species will be stopped within \SI{100}{\micro\metre}.
The total thickness of the collector sums up to \SI{10}{\milli\metre}.
The collector length is kept to a reasonable size of \SI{50}{\milli\metre} to ensure that it fits into the diagnostic ports of the \gls{gsi} linear accelerator UNILAC while keeping a safe distance to other diagnostics.
An SMA connector with a pin diameter of \SI{1.27}{\milli\metre} was employed to extract the signal from the FFC. The entire length of \SI{25}{\milli\metre} was utilized for tapering to the smaller diameter to achieve improved impedance matching.

The diameter of the collector is thicker than the outer diameter of the SMA connector, making it impossible to insert the collector through the openings for the SMA connectors in the chassis of the \gls{trcffc}. In the simulations, the surface currents are observed to flow directly from the center toward the connectors. The split-block technique \cite{Skaritka2022,Ruggiero2021}
is utilized, in which the chassis is divided into two halves along a plane perpendicular to the SMA ports and passing through the drift passage.
SMA connectors with loose inner pins are employed, allowing the chassis and the connector's insulator to be soldered to the collector prior to final assembly.
Since the SMA pins are loose and not fixed inside the connector, they can rotate, which may cause misalignment between the collector's drill hole and the pinhole of the heat shield.
This would compromise the geometrical \gls{se} suppression. While the collector is radially limited in its movement by the SMA pins, glass pins are added (see Fig.~\ref{fig-005}) to lock the rotation and ensure a well-defined longitudinal position of the collector inside the chassis.
Quartz glass was selected for the pins for its high heat resistance, low thermal expansion coefficients, mechanical stability, and the low permittivity $\epsilon_r\approx\num{3.75}$.

\begin{figure}[!htb]
    \centering
    \begin{subfigure}{0.49\textwidth}
        \includegraphics*[width=.95\columnwidth]{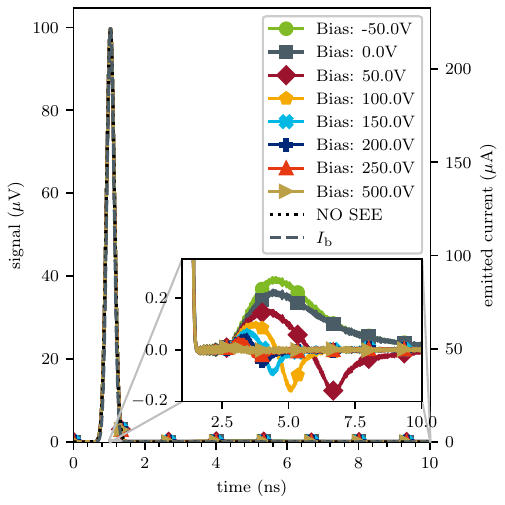}
        \caption{}
        \label{fig-007a}
    \end{subfigure}
    \hfill
    \begin{subfigure}{0.49\textwidth}
        \includegraphics*[width=.95\columnwidth]{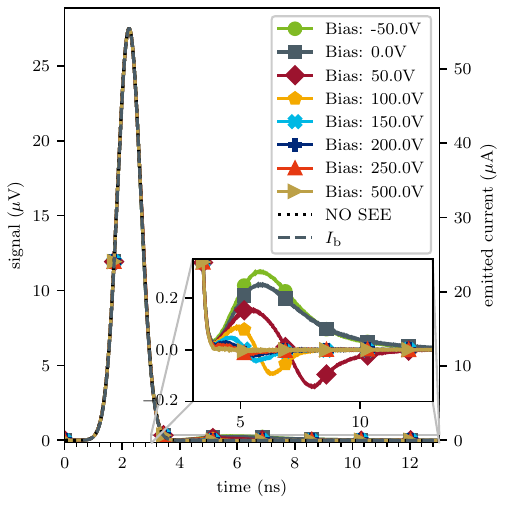}
        \caption{}
        \label{fig-007b}
    \end{subfigure}
    \caption{\gls{cst} PIC simulations of the \gls{trcffc} (see Fig.~\ref{fig-005}) with different bias settings and bunch lengths (\subref{fig-007a}) $\gls{sigmab}=\SI{100}{\pico\second}$ and (\subref{fig-007b}) $\gls{sigmab}=\SI{100}{\pico\second}$ with activated \gls{see}.}
    \label{fig-007}
\end{figure}

The effect on the \gls{se} suppression using the optimized \gls{trcffc} design is shown in Fig.~\ref{fig-007}. The same beam parameters are used as in the simulations for Figs.~\ref{fig-002} and \ref{fig-003}. Only \gls{se} through impinging ions on the collector are allowed.
No significant signal strength gain is observed despite the larger pinhole in the heat shield, because all emitted ions hit the collector in all simulations never touching any other surface before. Hence, the same number of ions is collected in each case.
In the new design, the geometrical \gls{se} suppression reduces the amplitude of the \gls{se} signal from \SI{12}{\micro\volt} (unbiased \gls{rcffc}) down to \SI{0.22}{\micro\volt}.
Additionally, the increased separation time minimizes the overlap between the ion and the \gls{se} signal down to \SI{300}{\pico\second}. The amplitude error due to the overlapping \gls{se} drops below \SI{1}{\micro\volt}.
The total impact of the \gls{se} on the ion signal is sufficiently low, making it likely that low-charge-state ion beams ($Z\leq10+$) will not require an additional bias scheme to suppress \gls{se}-induced signals to the noise level.

\subsection{Construction of 3D printed \gls{ffc}}
The key feature of the \gls{trcffc} is its deep drilled hole in the inner collector, which ensures high geometrical suppression of \gls{se}, as demonstrated in simulations (see \ref{Tapered_Radially_coupled_ffc_design_considerations_and_simulations}). As a consequence of this deep drill hole, the collector needs to be tapered down to the diameter of the inner pin of the connector used.
Accordingly, the most critical construction constraints are the taper angle of the collector, the cone in the chassis matching the \SI{50}{\ohm} geometry and the through holes for the glass pins for the alignment.
To evaluate the potential of additive manufacturing for beam diagnostics in heavy ion beam research, such as at \gls{gsi}, several collectors in slightly varying geometries were fabricated using additive manufacturing techniques. These were compared to conventionally machined collectors with respect to their RF-properties.
The collectors were printed at Fraunhofer-Institut für Werkstoff- und Strahltechnik IWS using Cu-ETP
with a purity of \SI{99.97}{\percent} copper in a TruPrint1000 Green Edition,
which operates with a green laser of \SI{515}{\nano\metre} and a maximal power of \SI{500}{\watt}. The collectors are printed vertically from port to port. There is no need for support structures because the taper angle is sufficiently small.
\begin{figure}[!htb]
    \centering
    \begin{subfigure}{0.49\textwidth}
        \includegraphics*[width=.9\columnwidth]{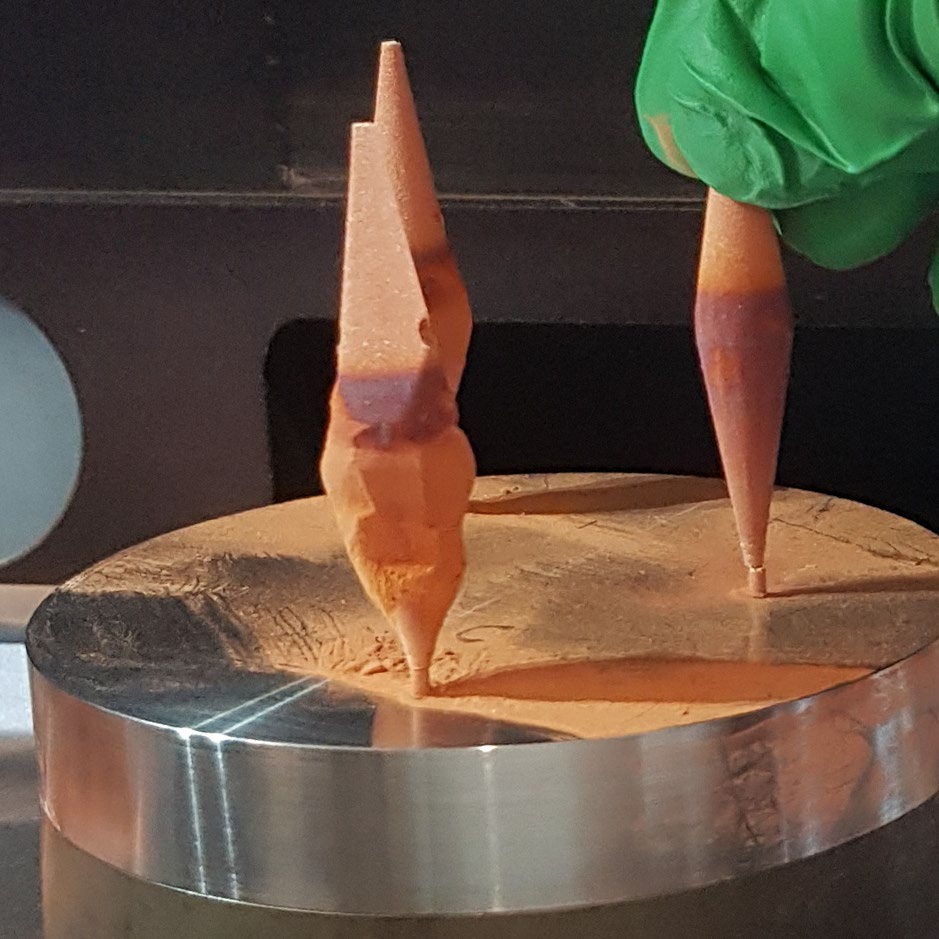}
        \caption{}
        \label{fig-008a}
    \end{subfigure}
    \hfill
    \begin{subfigure}{0.49\textwidth}
        \includegraphics*[width=.9\columnwidth]{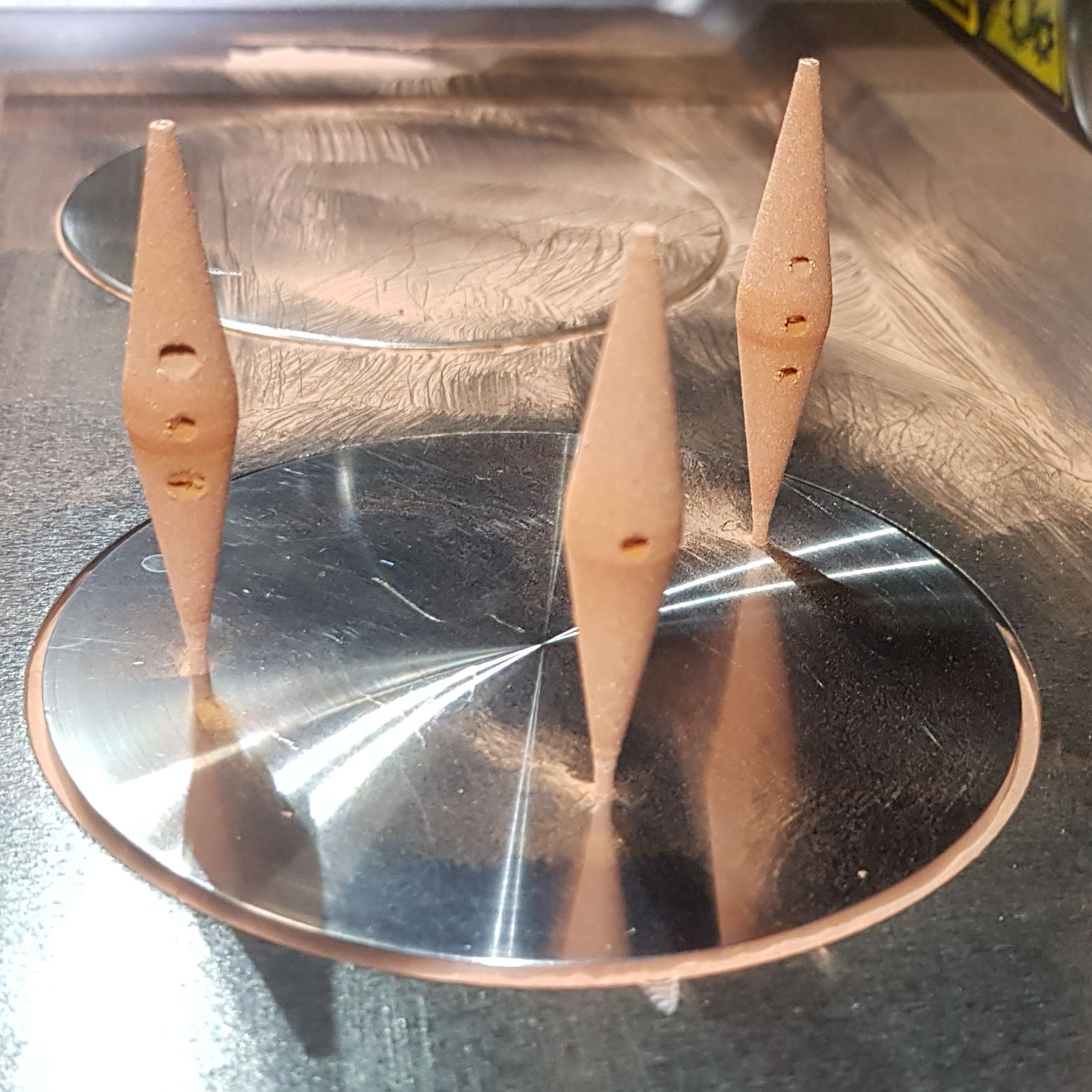}
        \caption{}
        \label{fig-008b}
    \end{subfigure}
    \caption{Printed collectors on the printing bed first batch (\subref{fig-008a}) and third batch (\subref{fig-008b}).}
    \label{fig-008}
\end{figure}
Figure~\ref{fig-008a} shows the first batch produced still attached on the build plate. This batch was printed without any waiting times, resulting in a strong coloration on the bottom side indicating overheating due to insufficient heat extraction through the build plate. The third batch (see Fig.~\ref{fig-008b}) has been printed with waiting times after each layer resulting in a \num{3} times higher build time. The decoloring is strongly reduced with only a small part closely below the collector hole being still decolored.
\begin{figure}[!htb]
    \centering
    \begin{subfigure}{0.49\textwidth}
        \includegraphics*[width=.9\columnwidth]{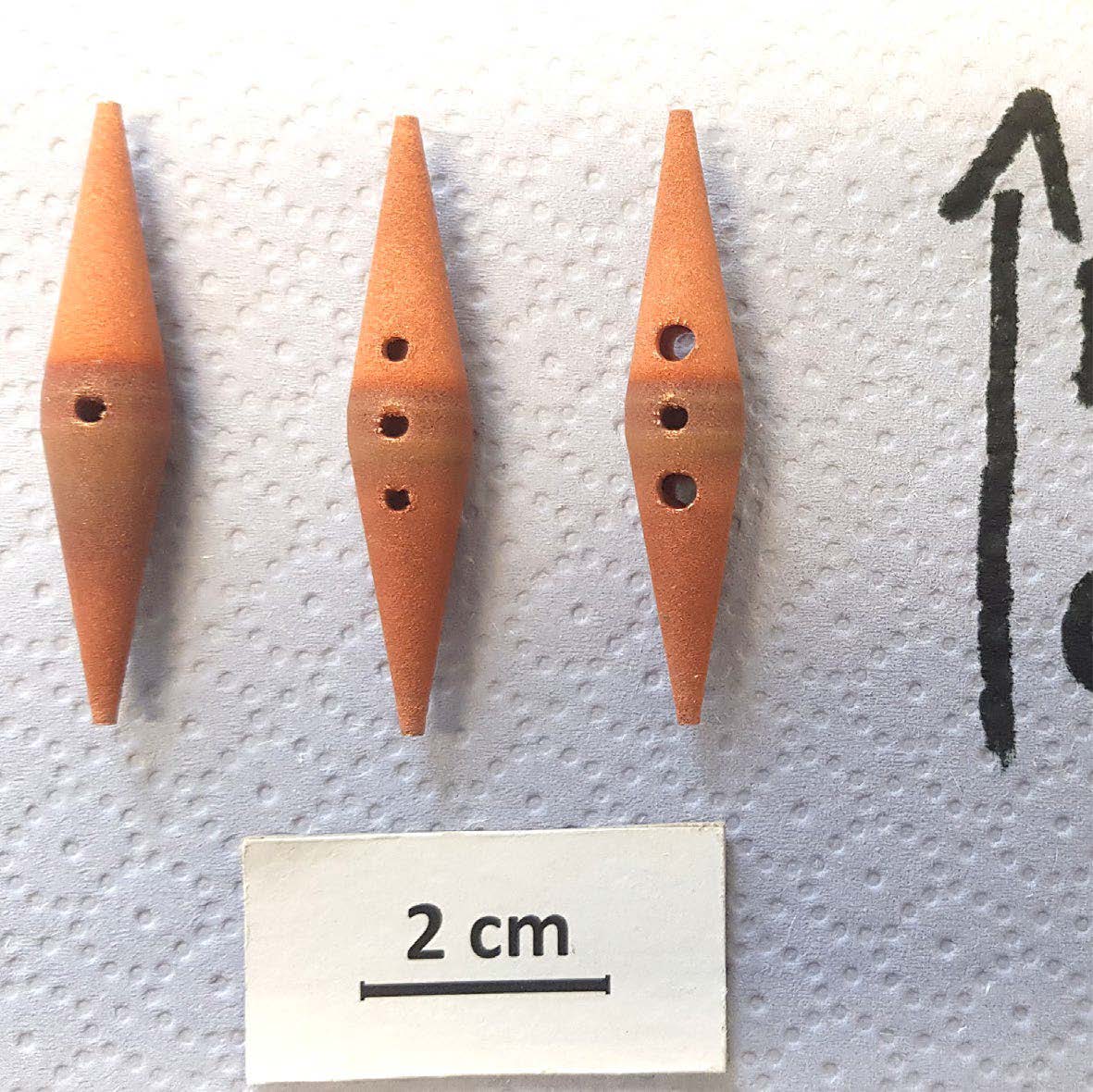}
        \caption{}
        \label{fig-009a}
    \end{subfigure}
    \hfill
    \begin{subfigure}{0.49\textwidth}
        \includegraphics*[width=.9\columnwidth]{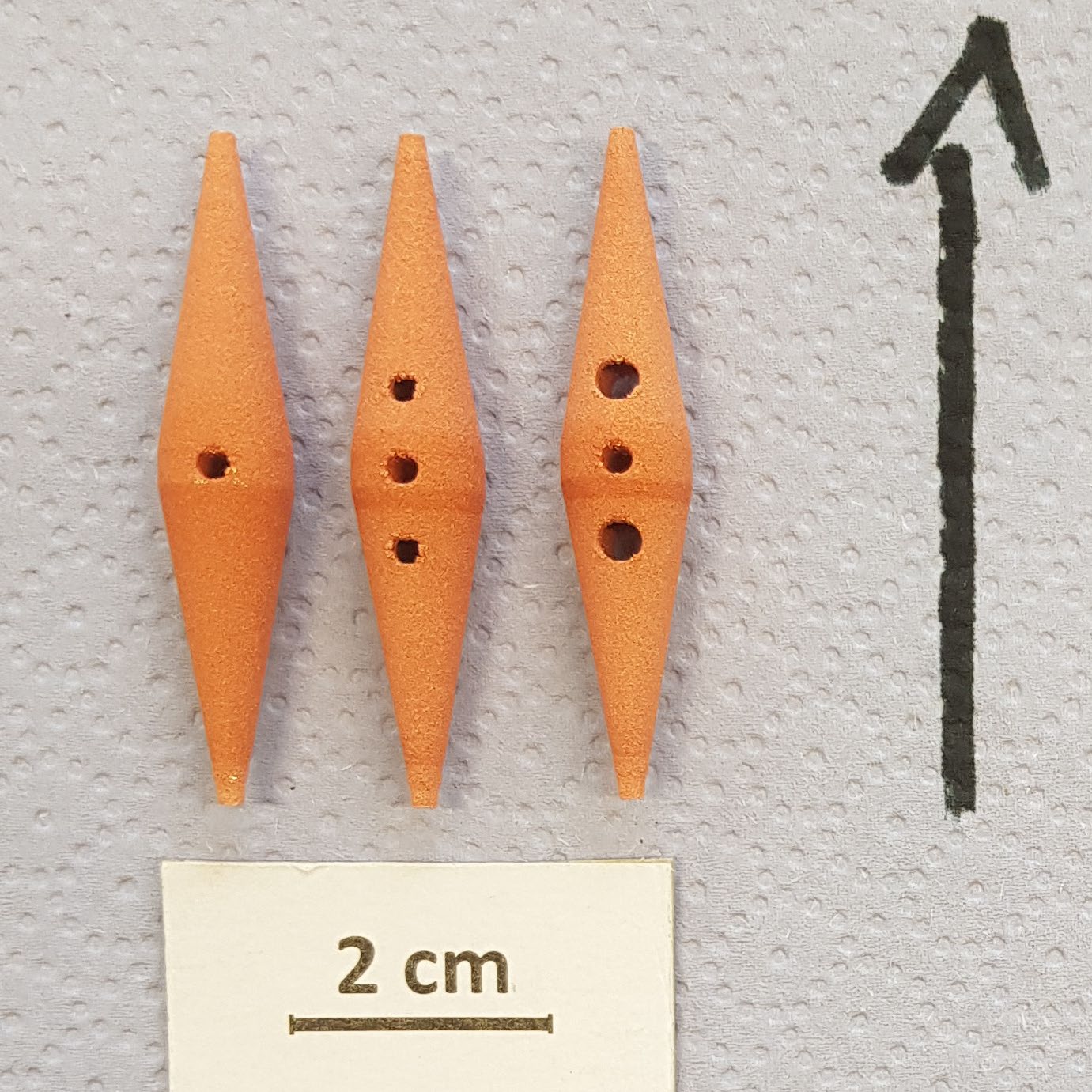}
        \caption{}
        \label{fig-009b}
    \end{subfigure}
    \caption{Collectors after cleaning from the powder, first batch (\subref{fig-009a}) and third batch (\subref{fig-009b}). The arrow marks the printing direction.}
    \label{fig-009}
\end{figure}

The collector hole and the holes for the glass pins are directly taken into account during the printing process. Since no support structures are used inside the drill holes, they become slightly oval. It is necessary to post-process the through holes of the alignment glass pins and the holes for the SMA pins. The printed copper material is relatively soft, allowing manual drilling.
A few fractals at the edges of the holes need to be removed as well. The surface has not been polished. The cleaned collectors may be seen in Fig.~\ref{fig-009}.
One of the very first designs was meant to utilize a flat tap connection SMA pin. This would have been advantageous preventing the rotating of the collector even without glass pins. However, the slit is too thin to be printed. Moreover, soldering the flat tab to the collector needs a different assembly approach than the split-block technique. Therefore, the round post  SMA connector was selected, with the disturbance introduced by the glass pins to the \SI{50}{\ohm} geometry being accepted.
There are three different versions of the collector printed in each batch: one without holes, one with \SI{2}{\milli\metre} and one with \SI{3}{\milli\metre} holes for glass pins. The different versions are used to study the influence of the glass on the RF-properties (see \ref{RF-characterization-of-the-rcffc-designs}).

\begin{figure}[h!]
    \centering
    \begin{subfigure}[b]{0.24\textwidth}
        \centering
        \includegraphics[width=\textwidth]{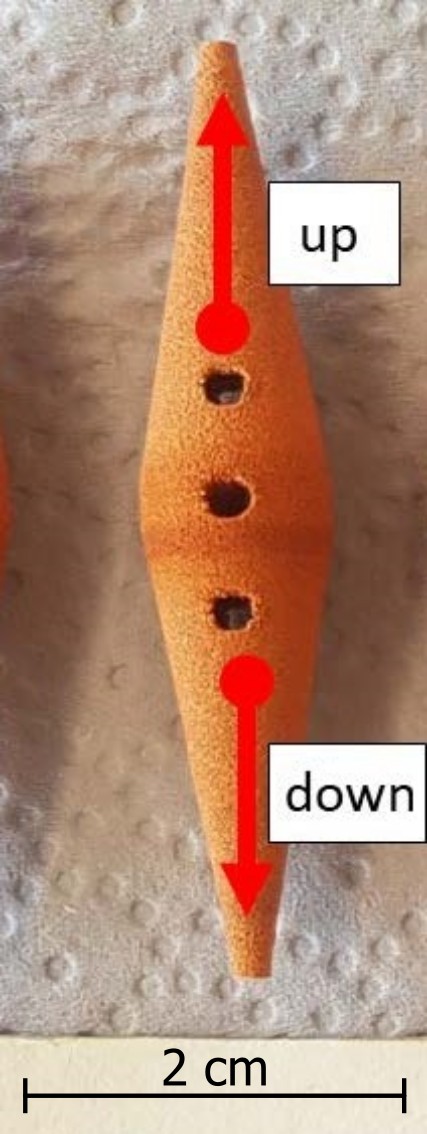}
        \caption{}
        \label{fig-010a}
    \end{subfigure}
    \hfill
    \begin{subfigure}[b]{0.73\textwidth}
        \centering
        \begin{subfigure}[b]{\textwidth}
            \centering
            \includegraphics[width=\textwidth]{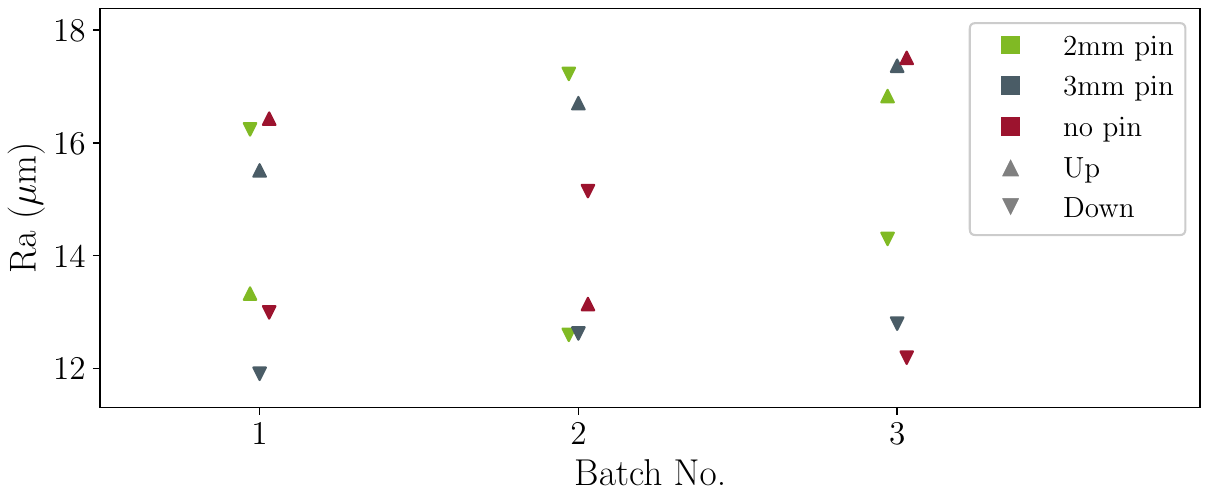}
            \caption{}
            \label{fig-010b}
        \end{subfigure}
        \vskip 0.5em 
        \begin{subfigure}[b]{\textwidth}
            \centering
            \includegraphics[width=\textwidth]{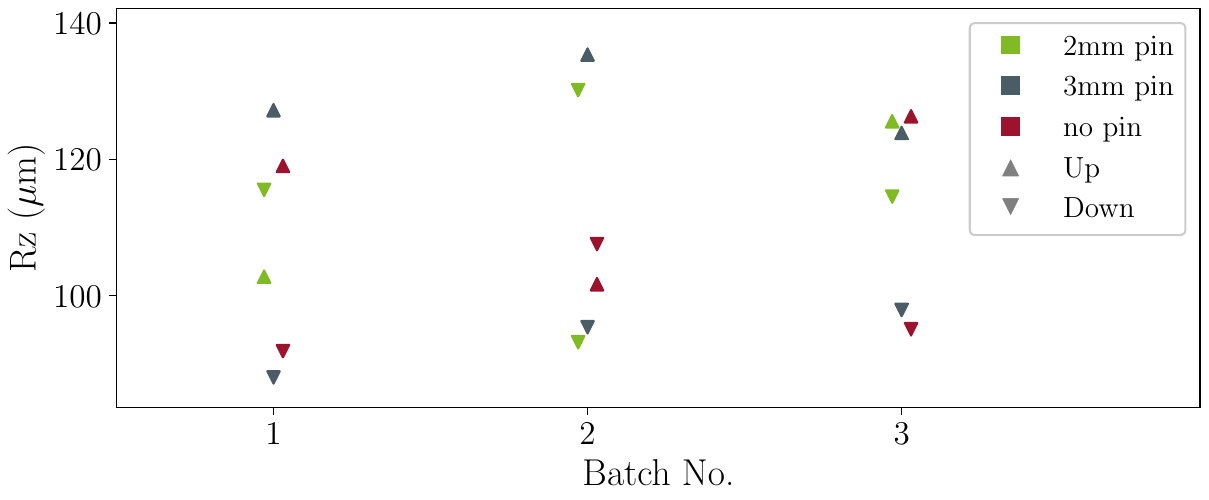}
            \caption{}
            \label{fig-010c}
        \end{subfigure}
    \end{subfigure}
    \caption{(\subref{fig-010a}) Measurement of the surface roughness starting from two points with a probe diameter of \SI{5}{\micro\metre} on a length of \SI{8}{\milli\metre} upwards and downwards, (\subref{fig-010b}) shows the measured average profile height deviations from the mean line $R_\text{a}$ and (\subref{fig-010c}) the maximum peak height to valley height $R_\text{z}$.}
    \label{fig-010}
\end{figure}

The surface roughness was measured using a Surfcom Touch 50 \cite{Accretech2024}. The probe diameter is \SI{5}{\micro\metre} and the measurement has been performed once for the upper cone and once for the lower cone of the collector for a distance of \SI{8}{\milli\metre}, as shown in Fig.~\ref{fig-010a}. The average profile height deviations from the mean line $h_\text{ref}$, $R_\text{a}$ (see Eq.~(\ref{eq-4})) \cite{DIN4768}, are shown in Fig.~\ref{fig-010a}.

\begin{linenomath}
    \begin{equation}
        R_\text{a} = \frac{1}{l} \int_0^l\left(\left|h(x)-h_\text{ref}\right|\right)dx
        \label{eq-4}
    \end{equation}
\end{linenomath}
The roughness ranges from \SIrange{12}{18}{\micro\metre}, with the topside being rougher than the lower part according to $R_\text{a}$. Also, the maximum peak height to valley height $R_\text{z}$ (see Eq.~(\ref{eq-5}), Fig.~\ref{fig-010c}) \cite{DIN4768} indicates higher roughness on the upper side. Overall, the roughness of the \SI{50}{\ohm} geometry varies by $\pm\SI{0.18}{\percent}$, which is negligible.

\begin{figure}[h!]
    \centering
    \begin{subfigure}[b]{0.45\textwidth}
        \centering
        \begin{subfigure}[b]{\textwidth}
            \centering
            \includegraphics[width=\textwidth]{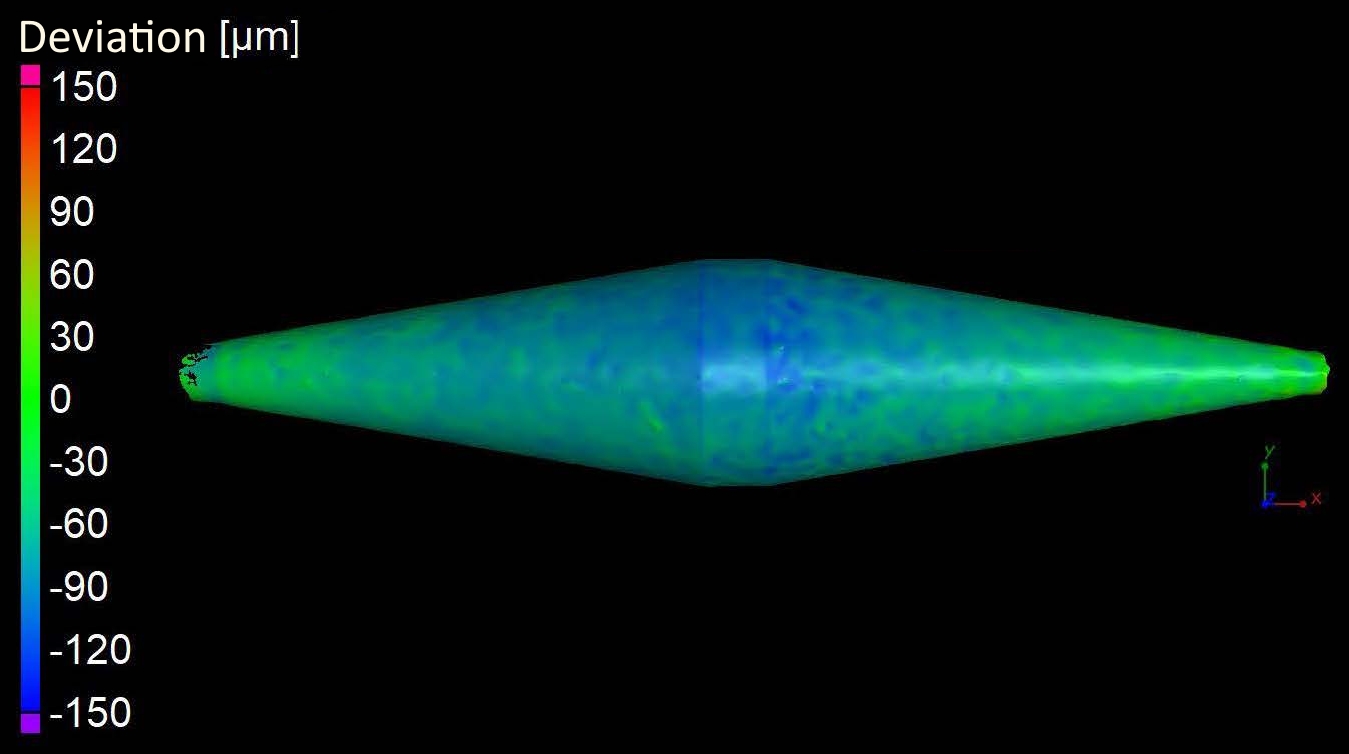}
            \caption{}
            \label{fig-011a}
        \end{subfigure}
        \vskip 0.5em 
        \begin{subfigure}[b]{\textwidth}
            \centering
            \includegraphics[width=\textwidth]{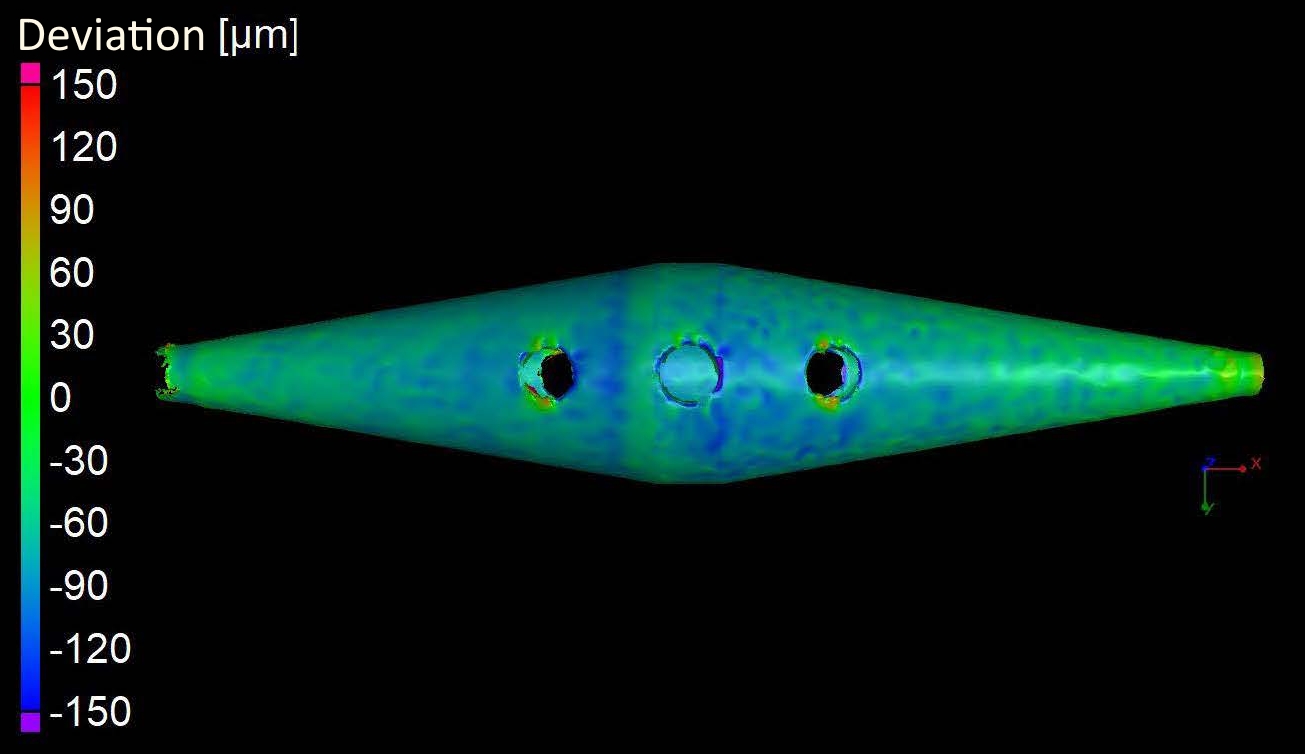}
            \caption{}
            \label{fig-011b}
        \end{subfigure}
        \vskip 0.5em 
        \begin{subfigure}[b]{\textwidth}
            \centering
            \includegraphics[width=\textwidth]{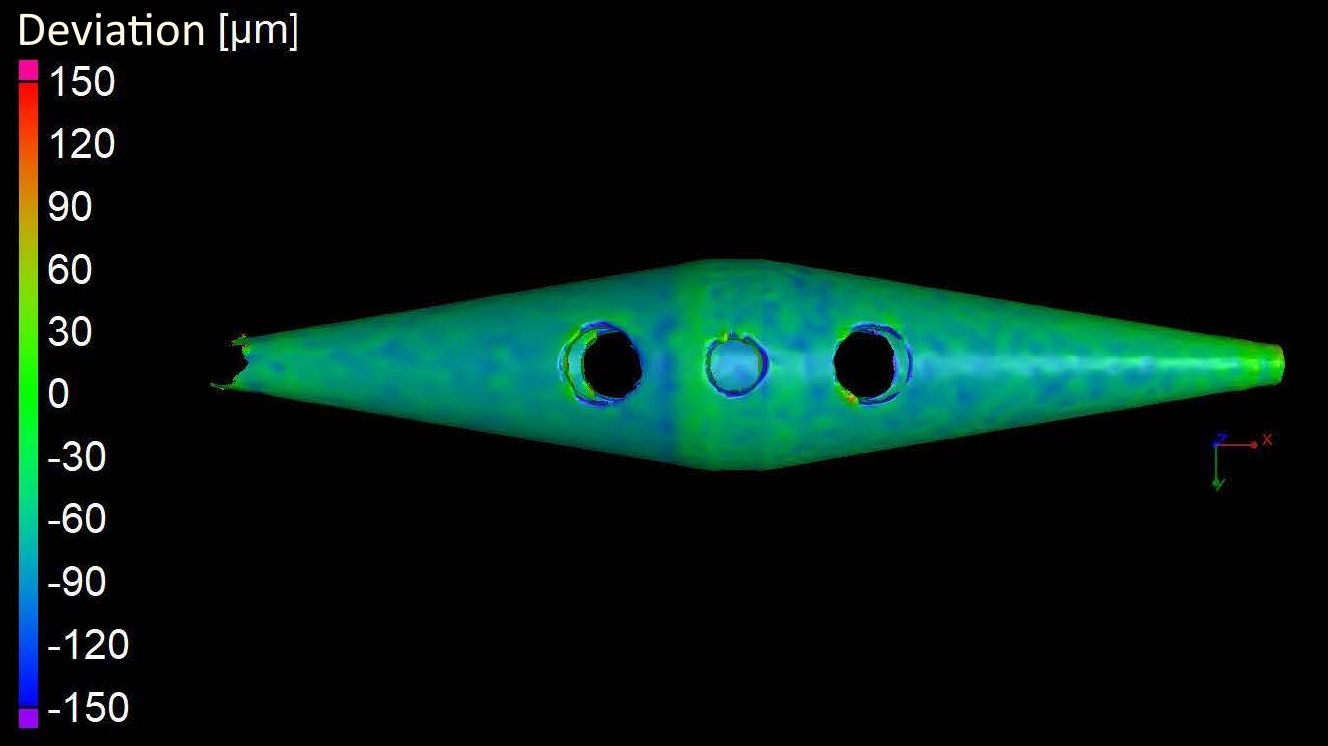}
            \caption{}
            \label{fig-011c}
        \end{subfigure}
    \end{subfigure}
    \hfill
    \begin{subfigure}[b]{0.45\textwidth}
        \centering
        \begin{subfigure}[b]{\textwidth}
            \centering
            \includegraphics[width=\textwidth]{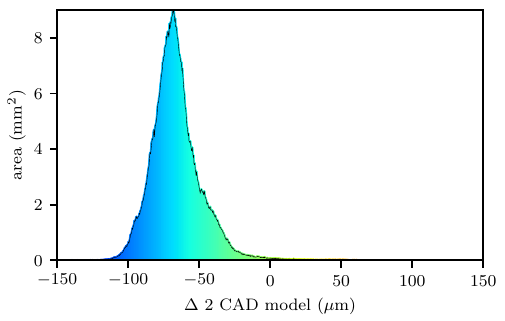}
        \end{subfigure}
        \vskip 0.50em 
        \begin{subfigure}[b]{\textwidth}
            \centering
            \includegraphics[width=\textwidth]{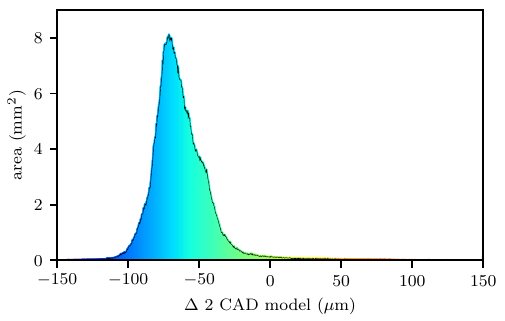}
        \end{subfigure}
        \vskip 1.25em 
        \begin{subfigure}[b]{\textwidth}
            \centering
            \includegraphics[width=\textwidth]{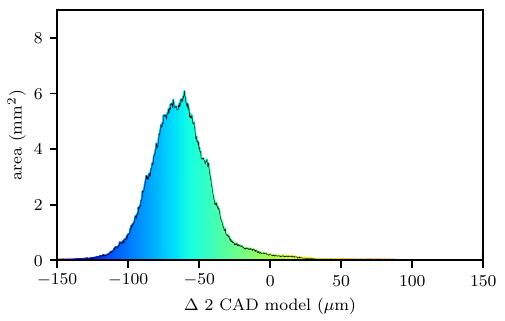}
        \end{subfigure}
        \vspace{-1.5em}
    \end{subfigure}
    \caption{3D scan of the printed collectors of batch 3 with color-coded difference to the CAD model for the geometries (\subref{fig-011a}) no glass pins, (\subref{fig-011b}) \SI{2}{\milli\metre} glass pins and (\subref{fig-011c}) \SI{3}{\milli\metre} glass pins.}
    \label{fig-011}
\end{figure}

\begin{linenomath}
    \begin{equation}
        \begin{split}
            R_\text{z}     & = \frac{\sum_{i=1}^n\left(r_{\text{p},i} + r_{\text{v},i}\right)}{n}                                        \\
            r_{\text{p},i} & = \max\left\{\left|h(x)-h_{\text{ref},i}\right|\right\} \left\vert \text{$x \in$ probe segment $i$}\right\} \\
            r_{\text{v},i} & = \max\left\{\left|h_{\text{ref},i}-h(x)\right|\right\} \left\vert \text{$x \in$ probe segment $i$}\right\}
        \end{split}
        \label{eq-5}
    \end{equation}
\end{linenomath}
The deviation from the CAD model was measured using a GOM ATOS Core 135
and is shown in Fig.~\ref{fig-011}. One challenge of additive manufacturing is the shrinkage of the body, which depends on the geometry and is difficult to predict. In our case, the shrinkage is about \SIrange{-20}{-100}{\micro\metre}, with the most common difference of \SI{-70}{\micro\metre}. This leads to an error of \SIrange{0.24}{1.2}{\percent} and most common \SI{0.84}{\percent}. In future iterations, the collector geometries are planned to be printed with radii increased by approximately \SI{50}{\micro\metre} to compensate for shrinkage, thereby reducing dimensional deviations and minimizing the impact on the \SI{50}{\ohm} geometry.
\clearpage

\subsection{RF-characterization of the radially coupled \gls{ffc} designs}
\label{RF-characterization-of-the-rcffc-designs}

\begin{figure}[h!]
    \centering
    \begin{subfigure}[b]{0.49\textwidth}
        \centering
        \includegraphics[width=\textwidth]{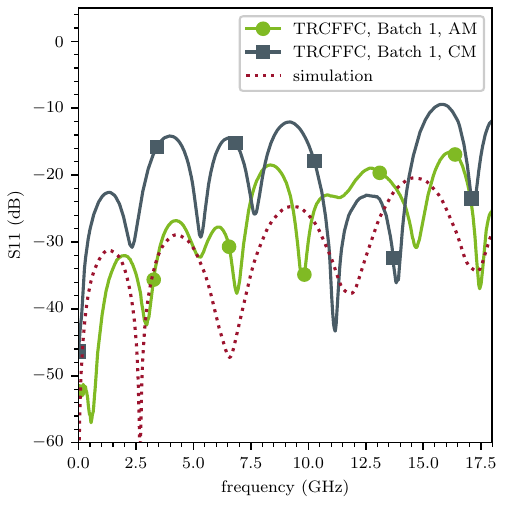}
        \caption{}
        \label{fig-012a}
    \end{subfigure}
    \hfill
    \begin{subfigure}[b]{0.49\textwidth}
        \centering
        \includegraphics[width=\textwidth]{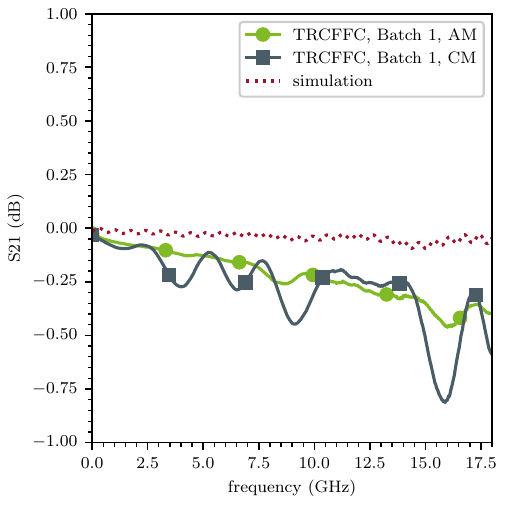}
        \caption{}
        \label{fig-012b}
    \end{subfigure}
    \caption{S-parameter (\subref{fig-012a}) S11 and (\subref{fig-012b}) S21
        measured using a \gls{trcffc} chassis with \SI{2}{\milli\metre} holes for glass pins and collectors without alignment holes. One collector was additively manufactured (AM)
        and the other conventionally machined (CM).}
    \label{fig-012}
\end{figure}

\begin{figure}[h!]
    \centering
    \begin{subfigure}[t]{0.65\textwidth}
        \centering
        \includegraphics[width=\textwidth]{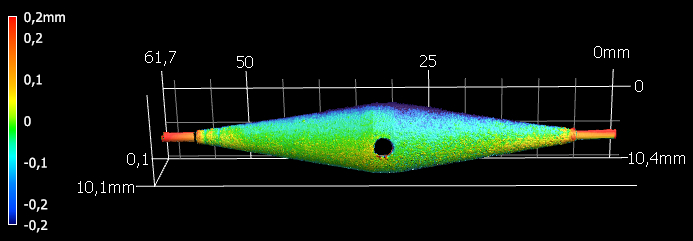}
        \caption{}
        \label{fig-013a}
    \end{subfigure}
    \hfill
    \begin{subfigure}[t]{0.30\textwidth}
        \centering
        \raisebox{-2.3em}{%
            \includegraphics[width=\textwidth]{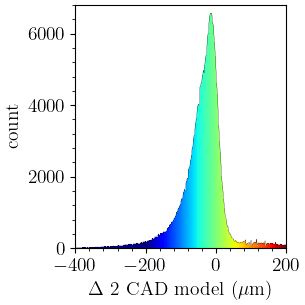}
        }
    \end{subfigure}

    \begin{subfigure}[b]{0.65\textwidth}
        \centering
        \raisebox{0.5em}{%
            \includegraphics[width=\textwidth]{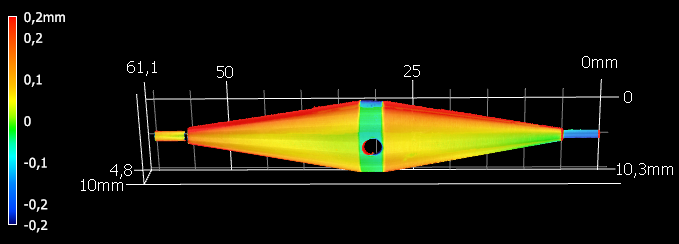}
        }
        \caption{}
        \label{fig-013b}
    \end{subfigure}
    \hfill
    \begin{subfigure}[b]{0.30\textwidth}
        \centering
        \raisebox{-2.3em}{%
            \includegraphics[width=\textwidth]{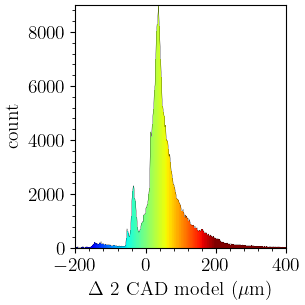}
        }
    \end{subfigure}

    \caption{The difference to the CAD model and its histogram is shown in (\subref{fig-013a}) for the \gls{AM} and in (\subref{fig-013b}) for the \gls{cm} version.}
    \label{fig-013}
\end{figure}

During the melting process in \gls{lpbf}, small defects can occur, leading to porosity formation. The conductivity $\sigma$ of the printed material is affected by this porosity, which adds up to other factors such as surface roughness and material impurity \cite{Robinson2022,Jadhav2021}. Additionally, the skin effect $\delta$ (Eq.~(\ref{eq-9})) is affected, which plays a significant role in high frequency signal losses within conductors.
\begin{linenomath}
    \begin{equation}
        \delta = \sqrt{\frac{1}{\pi f \mu_0 \mu_r \sigma}},
        \label{eq-9}
    \end{equation}
\end{linenomath}
with the frequency $f$, the free space permeability $\mu_0$, the relative permeability $\mu_r$ and the electrical conductivity $\sigma$.
For low power lasers (e.g. \SI{370}{\watt}), the conductivity of fused copper has been measured at approximately \SIrange{58}{73}{\percent}  of the \gls{iacs} \cite{Robinson2022}. When high power lasers ($>\SI{500}{\watt}$) are employed,  conductivities of up to \SI{94}{\percent} \cite{Jadhav2021} have been achieved. With a green laser even an electrical conductivity of 100 \gls{iacs} in the as-built condition is achievable \cite{Gruber2021}.
Therefore, the RF-properties of the \gls{trcffc} are analyzed using both a \gls{cm} and an \gls{AM} collector. The primary figure of merit for the RF-characterization are the S-parameters and the analysis of the temporal resolution, which was also compared to those obtained for the \gls{rcffc}.
In Figure~\ref{fig-012}, the measured S-parameters for a \gls{cm} and an \gls{AM} collector are shown in comparison with the \gls{cst} simulation results. These collectors have no holes for glass pins to study the differences in roughness and precision of the geometry apart from geometrically induced resonances. A FieldFox N9917A \cite{KeysightFieldFox2024} with \SI{0.5}{\metre} phase-stable SMA SUCOFLEX 126E cables \cite{HuberSuhner2018} was used. The calibration has been performed with a Keysight Ecal Module N7554A \cite{Keysight2024}.
It is observed that the reflections (Fig.~\ref{fig-012a})
are lower across nearly the entire \SI{18}{\giga\hertz} range, with the \gls{AM} collector performing better, on average, by about \SI{-7.5}{\decibel}. Additionally, for the transmission (Fig.~\ref{fig-012b}),
improved performance across the entire measurement range, with an average of \SI{0.05}{\decibel}.
The deviation from the CAD model was measured using a VR-6000 \cite{Keyence2024} for both the traditionally machined and additively manufactured version seen in Fig.~\ref{fig-013}. The measured profile were arranged to minimize the mean difference from the CAD model, resulting in smaller deviations compared to the previous full 3D scan in Fig.~\ref{fig-011}. The mean difference for the \gls{AM} version is \SI{13}{\micro\metre} and \SI{37}{\micro\metre} for the conventionally machined one. The surface roughness $R_\text{a}$ of \SI{2.4}{\micro\metre} of the \gls{cm} collector is significantly smoother, but still, the geometrical differences are stronger than the differences in the roughness. Hence, the performance difference with respect to the S-parameter is dominated by the geometrical differences rather than the surface roughness.

\begin{figure*}[!t]
    \centering
    \begin{subfigure}{0.49\textwidth}
        \includegraphics*[width=0.95\columnwidth]{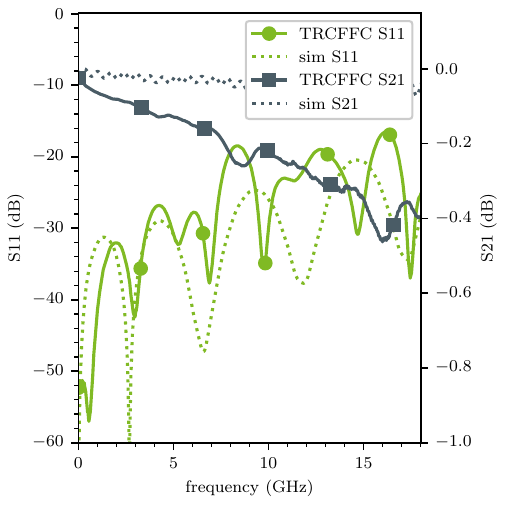}
        \caption{}
        \label{fig-014a}
    \end{subfigure}
    \hfill
    \begin{subfigure}{0.49\textwidth}
        \includegraphics*[width=0.95\columnwidth]{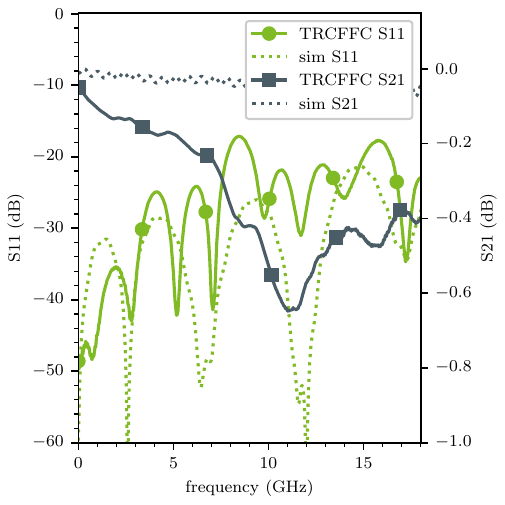}
        \caption{}
        \label{fig-014b}
    \end{subfigure}
    \hfill
    \begin{subfigure}{0.49\textwidth}
        \includegraphics*[width=0.95\columnwidth]{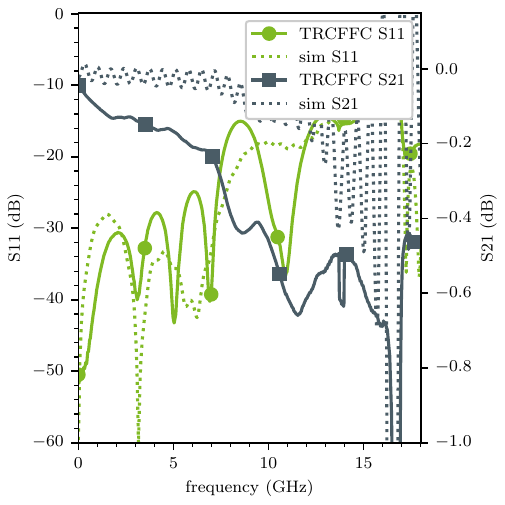}
        \caption{}
        \label{fig-014c}
    \end{subfigure}
    \hfill
    \begin{subfigure}{0.49\textwidth}
        \includegraphics*[width=0.95\columnwidth]{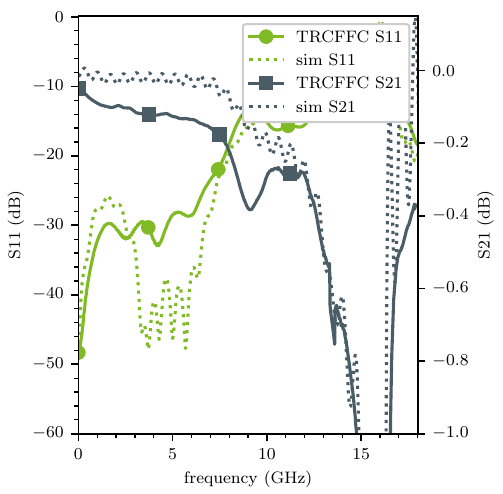}
        \caption{}
        \label{fig-014d}
    \end{subfigure}
    \caption{S-parameters of different designs. (\subref{fig-014a}) \gls{trcffc} with collector without holes for glass pins, (\subref{fig-014b}) \gls{trcffc} with collector with \SI{2}{\milli\metre} holes for glass pins, (\subref{fig-014c}) \gls{trcffc} with collector with inserted \SI{2}{\milli\metre} thick glass pins, (\subref{fig-014d}) \gls{trcffc} with collector with inserted \SI{3}{\milli\metre} thick glass pins.
    }
    \label{fig-014}
\end{figure*}

Subsequently, the focus will be set on the 3D-printed collectors to study the influence of the glass pins on the RF-properties, as the additively manufactured collector performed significantly better than the conventionally machined one.
Three different kinds of collectors were constructed, as shown in Fig.~\ref{fig-008}: one without, one with \SI{2}{\milli\metre} and one with \SI{3}{\milli\metre} holes for glass pins. The measurement results are compared with \gls{cst} simulations across the full \SI{18}{\giga\hertz} bandwidth of the SMA connectors (see Fig.~\ref{fig-014}). The best results are obtained without any glass nor holes for glass pins as seen in Fig.~\ref{fig-014a}. Adding \SI{2}{\milli\metre} diameter holes to the collector does not affect the S-parameters up to \SI{7.4}{\giga\hertz} as seen in Fig.~\ref{fig-014b}. However, the transmission is reduced above \SI{7.4}{\giga\hertz} to a minimum of \SI{-0.64}{\decibel} at \SI{11.2}{\giga\hertz} and the reflections increase up to \SI{-17}{\decibel}. Frequencies above \SI{7.4}{\giga\hertz} are therefore considered to be outside the bandwidth of the \gls{trcffc}.
After inserting \SI{2}{\milli\metre} glass pins into the alignment hole, the S-parameters remain close to the previous results up to \SI{7.4}{\giga\hertz} (see Fig.~\ref{fig-014c}). The transmission decreases for higher frequencies. The usage of the glass leads to two absorption peaks at \SI{14}{\giga\hertz} and \SI{16.8}{\giga\hertz}.
Increasing the diameter of the glass pins to \SI{3}{\milli\metre} (see Fig.~\ref{fig-014d}) levels the reflection to \SI{-30}{\decibel} till \SI{6}{\giga\hertz} and quickly increases to \SI{-12}{\decibel} afterwards. The absorption caused by the thicker glass pins drastically broadens the absorption peak at \SI{15.8}{\giga\hertz} with a transmission of only \SI{-7}{\decibel}. Therefore, the use of glass should be minimized to lower the disturbance of the \SI{50}{\ohm} geometry still stabilizing the collector position in the tapered coaxial design.
From the perspective of the S-parameters, this \gls{trcffc} design with \SI{2}{\milli\metre} glass pins could be used for bunches with a $\gls{sigmab}\geq\SI{135}{\pico\second}$.

Concerning the coaxial design, the \gls{trcffc} can be operated easily up to \SI{7}{\giga\hertz}, but the distance between the collector surface and the inner surface of the chassis also limits the usable bandwidth. The pre-field of the bunch advances ahead of the actual charges, prematurely reaching the collector and inducing a signal. This effect can be included into the calculation of the theoretical measurable bunch length \gls{sigmac} estimated by Eq.~(\ref{eq-1}) under the assumption that the diameter of the drill hole inside the collector and the drift tube after the heat shield is small compared to the distance $L$ between the collector and the inner chassis surface \cite{Mal2022}:
\begin{linenomath}
    \begin{equation}
        \gls{sigmac}^2 = \sigma_\text{b}^2 + \gls{sigmagap}^2 \,.
        \label{eq-1}
    \end{equation}
\end{linenomath}
where \gls{sigmagap} is the extension of the actual bunch length \gls{sigmab} and can be calculated from
\begin{linenomath}
    \begin{equation}
        \gls{sigmagap} = \kappa \cdot \frac{L}{\gls{beta} \cdot \gls{c0}} \,,
    \end{equation}
\end{linenomath}
with a geometrical constant $\kappa = 0.30$ \cite{FERMILABTM2641} and the velocity $\gls{beta} \gls{c0}$ of the bunch. To evaluate this effect, a series of simulations with very short bunches at different velocities was performed. The resulting simulated signal of these very short bunches is equivalent to the impulse response of the \glspl{ffc}.
In Table~\ref{tab-1} the resulting bunch length \gls{sigmas} of simulations for different bunch velocities \gls{beta} are shown next to the actual emitted bunch length \gls{sigmab} and the calculation of \gls{sigmac}. Thereby, we use the relation of the full width at half maximum $\text{FWHM} = \num{2.3548}\gls{sigmas}$ to derive the bunch length from the output signal of the \gls{trcffc}. Table~\ref{tab-2} shows the results for the same simulation settings for the \gls{rcffc}.
The generation of \gls{se} on the collector was activated to observe the effects of insufficient suppression, which leads to longer measured bunches. The bias voltage was set to \SI{0}{\volt}.
In both tables, it is evident that the time of flight of the particles through the gap between the chassis and the collector is the dominant parameter for very short pulses (\SI{10}{\pico\second}). The higher \gls{beta}, the lower is the impact of \gls{sigmagap} and hence, the precision of the \gls{ffc} rises. In comparison to the \gls{rcffc}, the temporal resolution of the \gls{trcffc} should be lower by a factor of \num{1.76} w.r.t. the theoretical model $\sigma_c$. Examining the simulation results, the factor between the temporal resolution of both designs is estimated to a factor of
\num{2.5}. Increasing the diameter of the collector for the \gls{trcffc} to improve the \gls{se} suppression results in a trade-off, as it comes at the cost of a lower bandwidth.
However, as the bunch length increases, the deviation in the \gls{trcffc} simulations for estimating the bunch length \gls{sigmas} at $\gls{beta}=\SI{15}{\percent}$ becomes less than \SI{10}{\percent} for bunch widths \gls{sigmab} exceeding $\gls{sigmab}=\SI{100}{\pico\second}$. For \SI{400}{\pico\second}, the deviation remains below \SI{1}{\percent} even without bias potential.
The higher bandwidth of the \gls{rcffc} geometry results in a higher temporal resolution also for shorter bunches resulting in a lower deviation, e.g. \SI{1.7}{\percent} for $\gls{sigmab}=\SI{100}{\pico\second}$ at $\gls{beta}=\SI{15}{\percent}$. Without an appropriate bias scheme, bunch shapes of $\gls{sigmab}\geq\SI{400}{\pico\second}$ are elongated. The lower temporal separation \gls{tsep} of the \gls{rcffc} leads to an overlapping of the electron and ion signal resulting in an deviation of \SI{17.5}{\percent} from the actual simulated bunch length.
The finite temporal resolution at speed of light is \SI{2.9}{\pico\second} for the \gls{rcffc} and \SI{5.3}{\pico\second} for the \gls{trcffc}.

\begin{table}[H]
    \caption{Timings \gls{trcffc}.\label{tab1}}
    \begin{tabularx}{\textwidth}{CCCCC}
        \toprule
        \textbf{beam velocity $\gls{beta}$} (\si{\percent}) & \textbf{$\sigma_\text{b}$} (ps) & \textbf{\gls{sigmac}} (ps) & \textbf{\gls{sigmas}} (ps) & \textbf{temporal resolution} (ps) \\
        \midrule
        1                                                   & 10                              & 650                        & 746.4                      & 745.3                             \\
        5                                                   & 10                              & 130                        & 146.9                      & 146.6                             \\
        15                                                  & 10                              & 44.5                       & 50.8                       & 49.8                              \\
        15                                                  & 100                             & 109.1                      & 110.7                      & 47.5                              \\
        15                                                  & 400                             & 402.4                      & 403.4                      & 52.3                              \\
        25                                                  & 10                              & 26                         & 32.8                       & 31.2                              \\
        \bottomrule
    \end{tabularx}
    \label{tab-1}
\end{table}

\begin{table}[H]
    \caption{Timings \gls{rcffc}.\label{tab2}}
    \begin{tabularx}{\textwidth}{CCCCC}
        \toprule
        \textbf{beam velocity $\gls{beta}$} (\si{\percent}) & \textbf{$\sigma_\text{b}$} (ps) & \textbf{\gls{sigmac}} (ps) & \textbf{\gls{sigmas}} (ps) & \textbf{temporal resolution} (ps) \\
        \midrule

        1                                                   & 10                              & 370                        & 300.8                      & 300.6                             \\
        5                                                   & 10                              & 74.1                       & 60.9                       & 60.1                              \\
        15                                                  & 10                              & 26.7                       & 21.8                       & 19.4                              \\
        15                                                  & 100                             & 103                        & 101.7                      & 18.5                              \\
        15                                                  & 400                             & 400.8                      & 470.9                      & 248.5                             \\ 
        25                                                  & 10                              & 14.8                       & 15                         & 11.2                              \\
        \bottomrule
    \end{tabularx}
    \label{tab-2}
\end{table}

\begin{figure*}[!t]
    \centering
    \begin{subfigure}{0.45\textwidth}
        \includegraphics*[width=0.95\columnwidth]{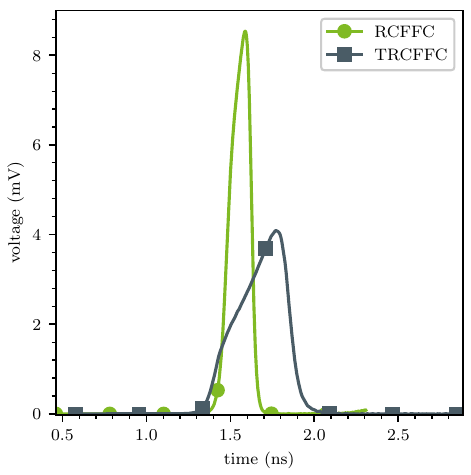}
        \caption{}
        \label{fig-015a}
    \end{subfigure}
    \hfill
    \begin{subfigure}{0.45\textwidth}
        \includegraphics*[width=0.95\columnwidth]{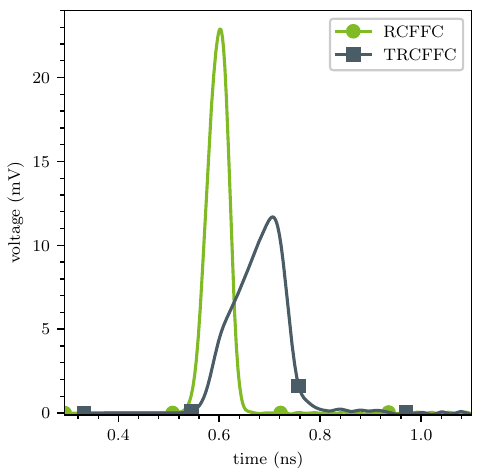}
        \caption{}
        \label{fig-015b}
    \end{subfigure}
    \hfill
    \begin{subfigure}{0.45\textwidth}
        \includegraphics*[width=0.95\columnwidth]{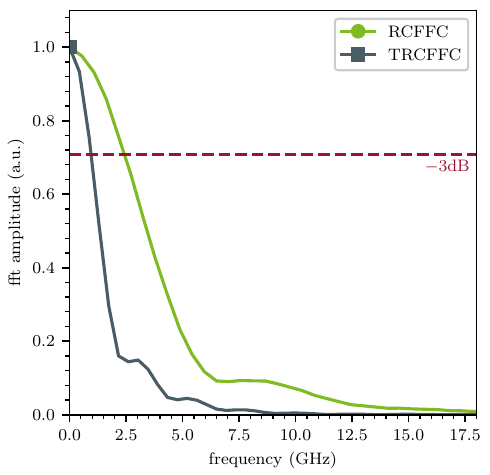}
        \caption{}
        \label{fig-015c}
    \end{subfigure}
    \hfill
    \begin{subfigure}{0.45\textwidth}
        \includegraphics*[width=0.95\columnwidth]{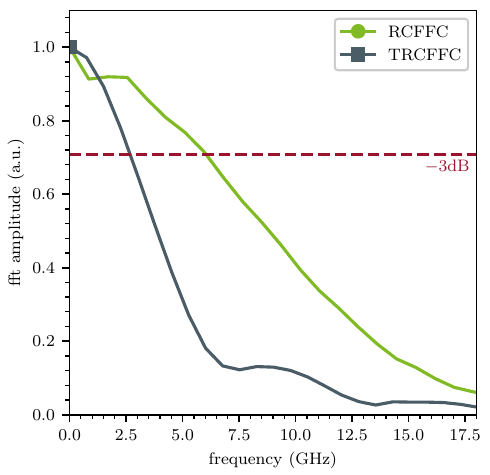}
        \caption{}
        \label{fig-015d}
    \end{subfigure}
    \caption{\gls{cst} HFTD simulation of an $\gls{sigmab}=\SI{10}{\pico\second}$ for velocity $\gls{beta} \gls{c0}$ (\subref{fig-015a}) \SI{5}{\percent} and (\subref{fig-015b}) \SI{15}{\percent} for the geometries \gls{rcffc} and \gls{trcffc} in comparison and respectively the FFT of the simulated signal for (\subref{fig-015c}) \SI{5}{\percent} and (\subref{fig-015d}) \SI{15}{\percent} indicating the \SI{-3}{\decibel} bandwidth as dashed line.}
    \label{fig-015}
\end{figure*}

A second way to look at the temporal resolution is to estimate the \SI{-3}{\decibel} bandwidth of the \glspl{ffc}. First, an FFT of the simulated signals is performed, as shown in Fig.~\ref{fig-015}, followed by searching for the intersection of the amplitude signal with the level \num{0.707}, which is equivalent to \SI{-3}{\decibel} degradation in signal power. In the case of $\gls{beta}=\SI{5}{\percent}$ (see Fig.~\ref{fig-015c}) the bandwidth of the \gls{rcffc} design is \SI{2.432}{\giga\hertz} (\gls{trcffc} \SI{0.953}{\giga\hertz}) and in the case of $\gls{beta}=\SI{15}{\percent}$ (see Fig.~\ref{fig-015d}), the bandwidths are \SI{6.064}{\giga\hertz} and \SI{2.685}{\giga\hertz} respectively, which corresponds to a factor of \numrange{2.5}{2.25} between the bandwidths of both devices. This is in line with the evaluation of the FWHM width estimation precision for very short bunches of \num{2.45}, as discussed earlier in this section.

\clearpage

\subsection{Measurements}

\begin{figure}[!htb]
    \centering
    \vspace{0.25em}
    \begin{subfigure}{0.49\textwidth}
        \includegraphics*[width=0.95\columnwidth]{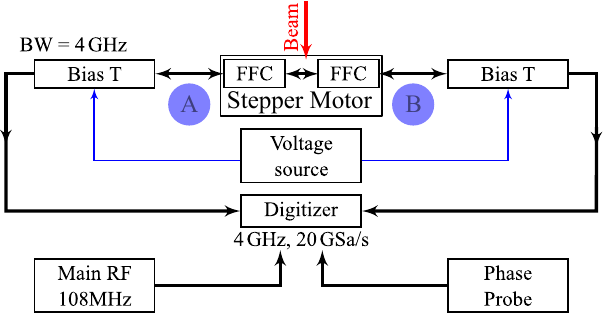}
        \caption{}
        \label{fig-016a}
    \end{subfigure}
    \hfill
    \begin{subfigure}{0.49\textwidth}
        \includegraphics*[width=0.95\columnwidth]{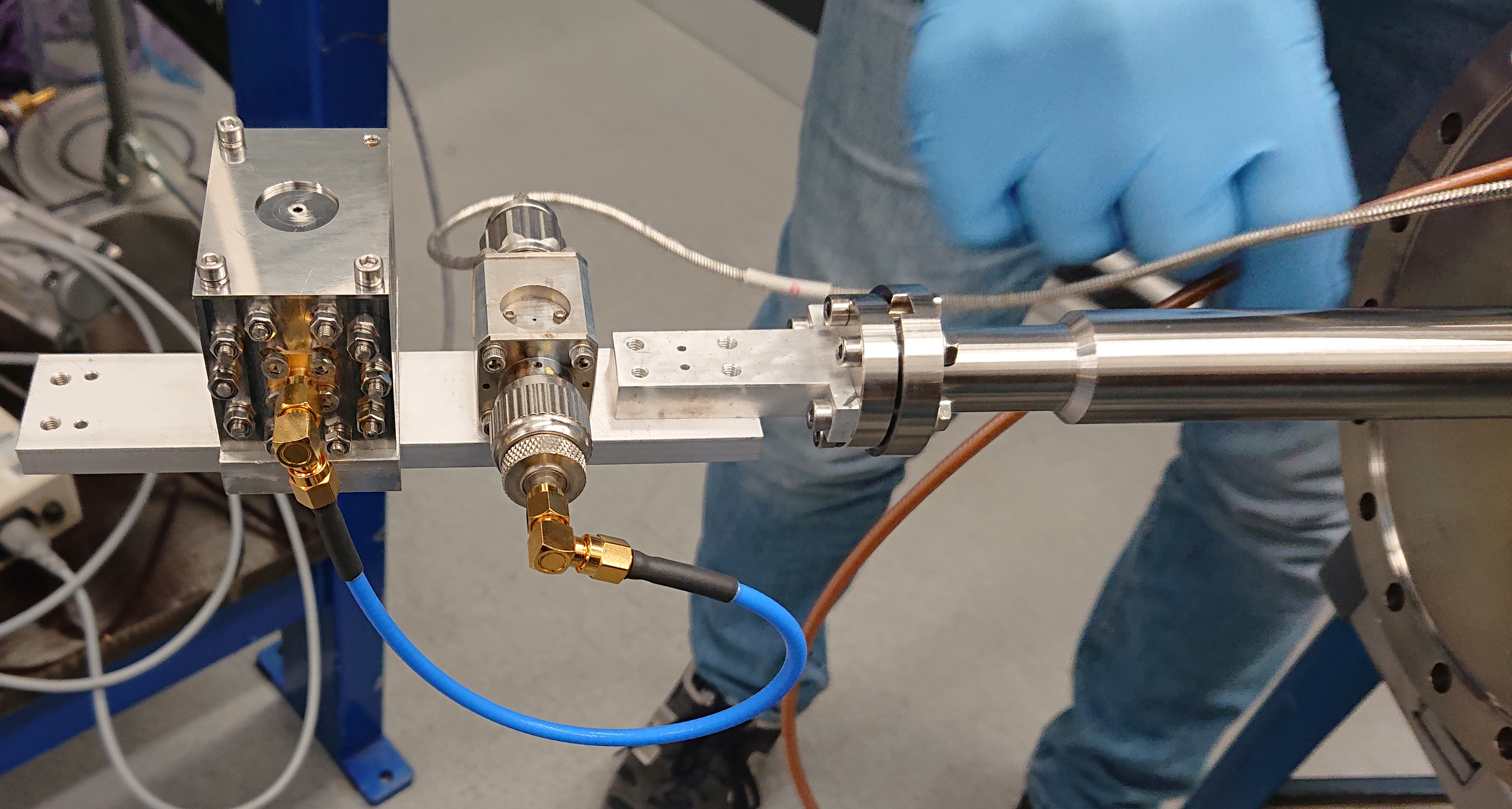}
        \caption{}
        \label{fig-016b}
    \end{subfigure}
    \hfill
    \vspace{0.25em}
    \begin{subfigure}{0.49\textwidth}
        \includegraphics*[width=0.95\columnwidth]{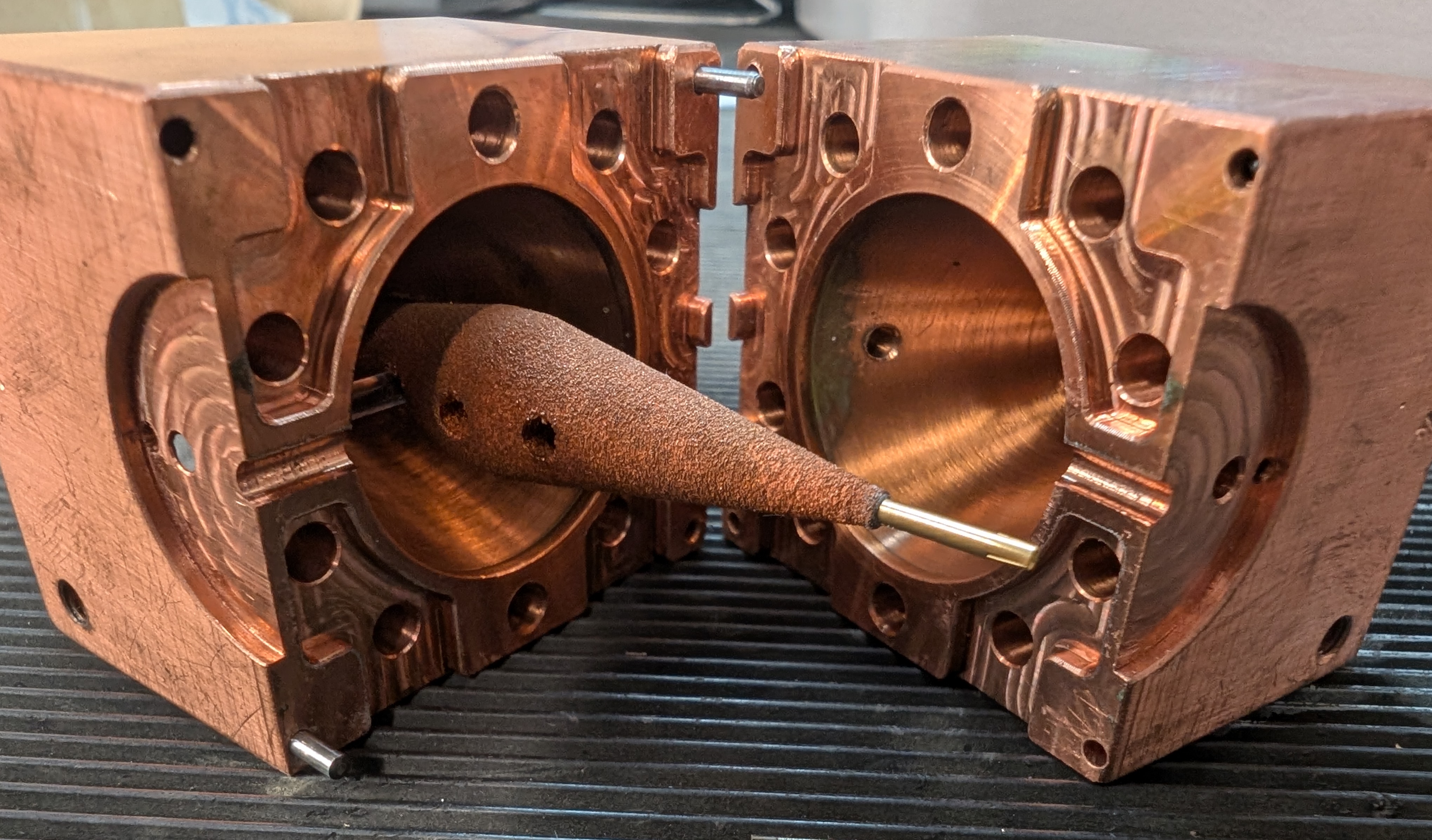}
        \caption{}
        \label{fig-016c}
    \end{subfigure}
    \hfill
    \begin{subfigure}{0.49\textwidth}
        \includegraphics*[width=0.95\columnwidth]{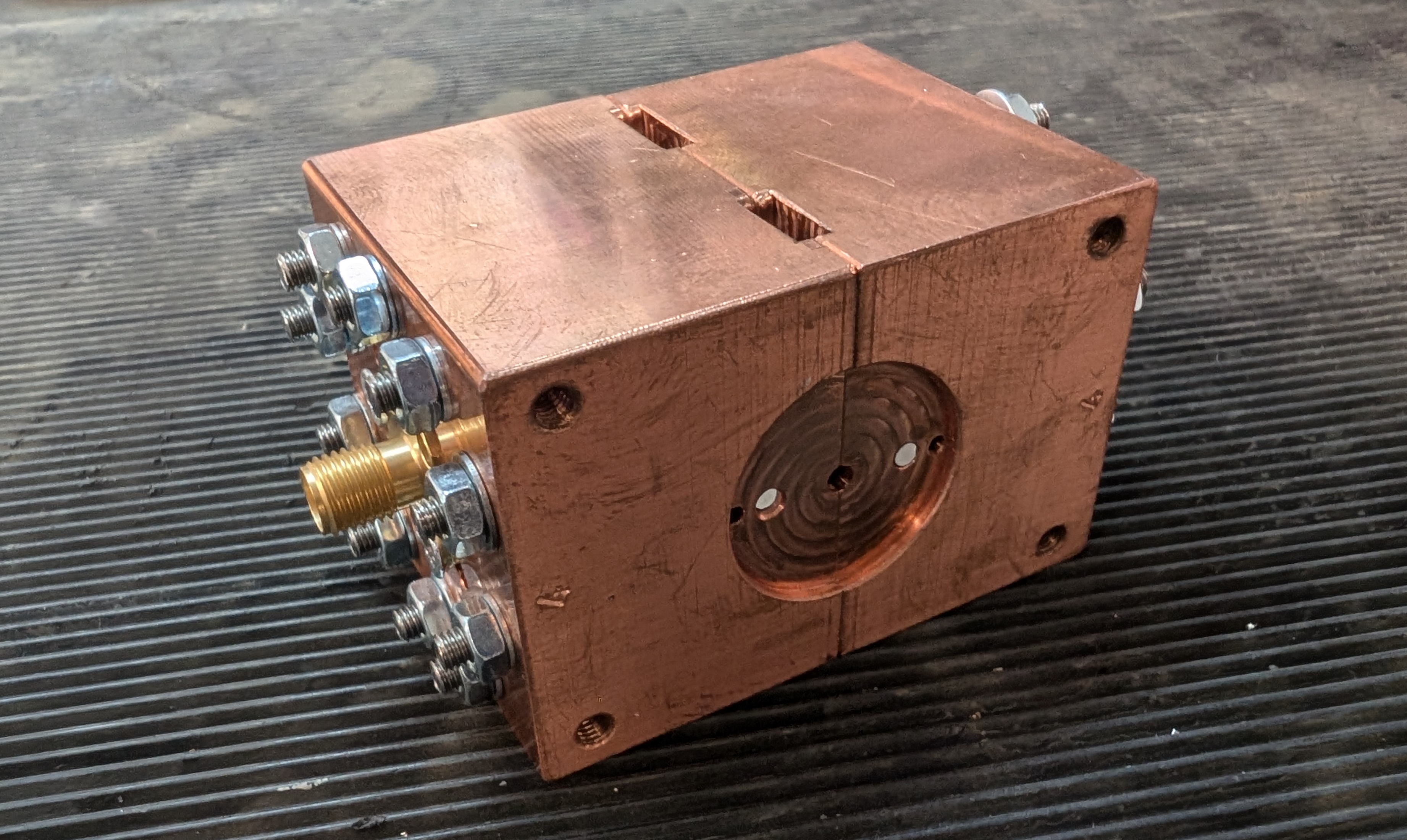}
        \caption{}
        \label{fig-016d}
    \end{subfigure}
    \caption{Experimental setup in the \gls{gsi} experimental cave X2: (\subref{fig-016a}) scheme of the signal paths and (\subref{fig-016b}) installation of the \gls{rcffc} and the \gls{trcffc} on a stepper motor. Assembly of a full copper version of the \gls{trcffc} with open chassis (\subref{fig-016c}) demonstrating the positioning of the collector using a glass pin and the closed chassis (\subref{fig-016d}).}
    \label{fig-016}
\end{figure}

\begin{figure*}[!t]
    \centering
    \begin{subfigure}{0.49\textwidth}
        \includegraphics*[width=0.95\columnwidth]{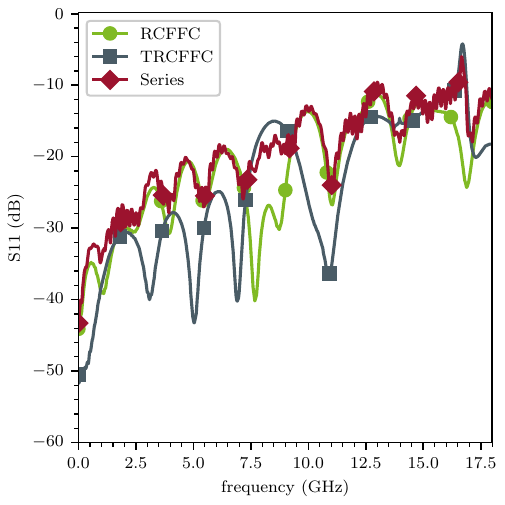}
        \caption{}
        \label{fig-017a}
    \end{subfigure}
    \hfill
    \begin{subfigure}{0.49\textwidth}
        \includegraphics*[width=0.94\columnwidth]{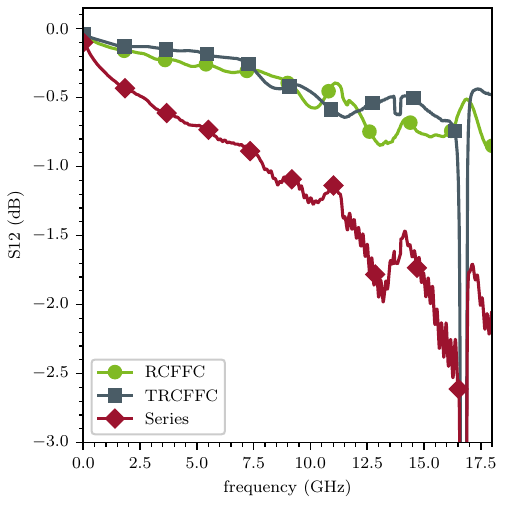}
        \caption{}
        \label{fig-017b}
    \end{subfigure}
    \caption{Measured S11 (\subref{fig-017a}) and S21 (\subref{fig-017b}) of a \gls{rcffc} and a \gls{trcffc} seperately and installed in a series circuit.
    }
    \label{fig-017}
\end{figure*}

A measurement campaign was performed to compare the two radially coupled \glspl{ffc} discussed above at the X2 measurement station at \gls{gsi}. The measurements were performed with an \gls{ar10+} beam at \SI{11.4}{\mega\electronvolt\per u}. Since there were only two vacuum feedthroughs available, \gls{rcffc} and \gls{trcffc} were connected in series onto a stepper motor (see Fig.~\ref{fig-016b}). Prior to the beam measurements, the RF-properties were examined in terms of reflection and transmission of each single device and connected in series as shown in Fig.~\ref{fig-017}.
The used \gls{rcffc} and \gls{trcffc} perform similar, so that the combined reflection properties remain below \SI{-20}{\decibel} till \SI{7}{\giga\hertz}. The transmission is dominated by the additional \SI{25}{\centi\metre} long SUCOFLEX 126E cable between them down to \SI{-0.8}{\decibel} till \SI{7}{\giga\hertz}. Using a shorter cable would further improve the transmission.
So in each case, the other \gls{ffc} behaves like a coaxial cable and does not affect the measurements.
The position of the \glspl{ffc} was alterable in the horizontal plane by the stepper motor. The beam irradiated only one of the two \glspl{ffc} at a given time. A measurement scheme with all the components is given in Fig.~\ref{fig-016a}. The bias voltage was applied through Mini-Circuits ZX85-12G-S+ bias Tees \cite{MiniCircuitsBiasTee2024} from both ports, ensuring that no DC current through the \glspl{ffc} is induced. The signal is amplified with a broadband (low-noise) amplifier Mini-Circuits ZX60-14LN-S+ \cite{MiniCircuitsAmplifier2024} and measured with a four-channel Lecroy WaveRunner 9404 \cite{TeledyneLeCroy2024} oscilloscope with \SI{4}{\giga\hertz} bandwidth and \SI{20}{\giga Sa \per\second}. In addition, the master RF signal of the UNILAC cavities was recorded to evaluate the bunch movements with respect to RF. The signal from the phase probe (PP) installed $\approx\SI{0.9}{\metre}$ in front of the \glspl{ffc} was also available for time of arrival triggering. An open, all copper version of the \gls{trcffc} can be seen in Fig.~\ref{fig-016c}. The collector is arranged with a glass pin inside the chassis. The sealing RF-surfaces are shown.

\begin{figure}[!htb]
    \centering
    \begin{subfigure}{0.49\textwidth}
        \includegraphics*[width=0.95\columnwidth]{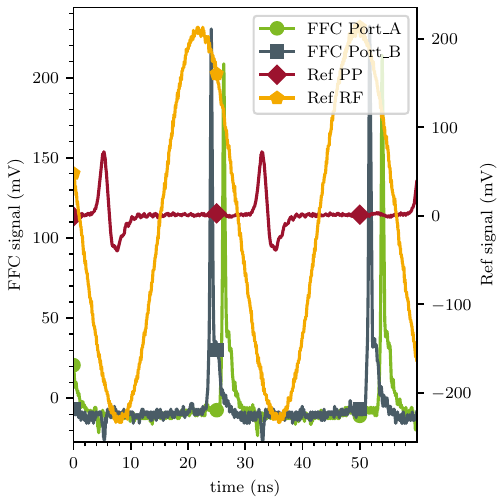}
        \caption{}
        \label{fig-018a}
    \end{subfigure}
    \hfill
    \begin{subfigure}{0.49\textwidth}
        \includegraphics*[width=0.95\columnwidth]{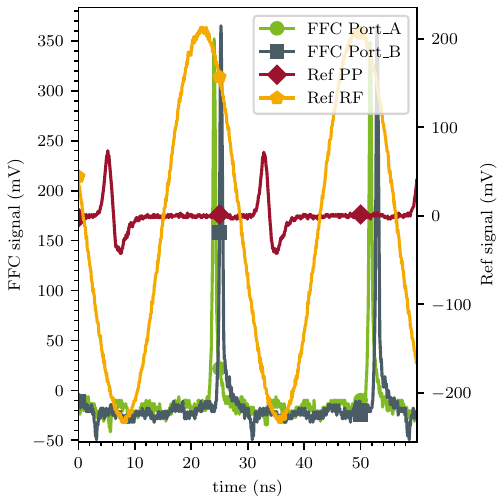}
        \caption{}
        \label{fig-018b}
    \end{subfigure}
    \caption{Two example measurements showing the phase probe and RF-reference signal together with the measured signals of port A and B using (\subref{fig-018a}) \gls{rcffc} and (\subref{fig-018b}) \gls{trcffc} as measurement device with an \gls{ar10+} beam at \SI{11.4}{\mega\electronvolt\per u}.}
    \label{fig-018}
\end{figure}

In Figures~\ref{fig-018a} and \ref{fig-018b}, examples of two consecutive bunches within a macropulse measured with the \gls{rcffc} (left) and the \gls{trcffc} (right) at the same position w.r.t. the beam axis without biasing are shown. The phase probe signal and reference RF are also marked. The time difference between the \glspl{ffc} A and B ports is a result of different signal path lengths. \
The difference in the amplitude of the signals arises mainly due to the different damping from the additional cable between the \glspl{ffc} and a small additional reflection of the second \gls{ffc}.
Inline with the expectations, the amplitude of the ion peak of the \gls{trcffc} is higher by a factor of \num{1.6} compared to the \gls{rcffc}. Furthermore, the tail of the \gls{se} is lower and shorter in the \gls{trcffc} due to the stronger geometrical \gls{se} suppression.
\begin{figure}[!htb]
    \centering
    \begin{subfigure}{0.49\textwidth}
        \includegraphics*[width=.95\columnwidth]{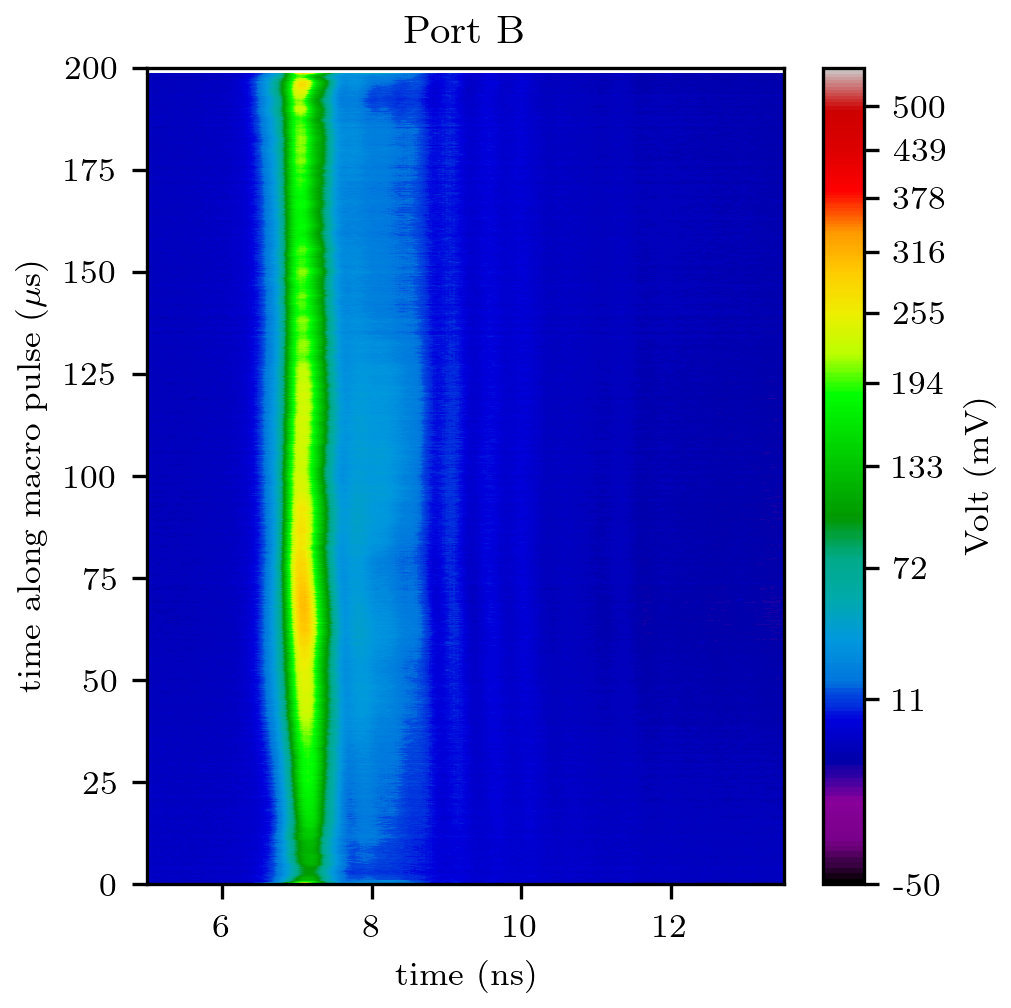}
        \caption{}
        \label{fig-019a}
    \end{subfigure}
    \hfill
    \begin{subfigure}{0.49\textwidth}
        \includegraphics*[width=.95\columnwidth]{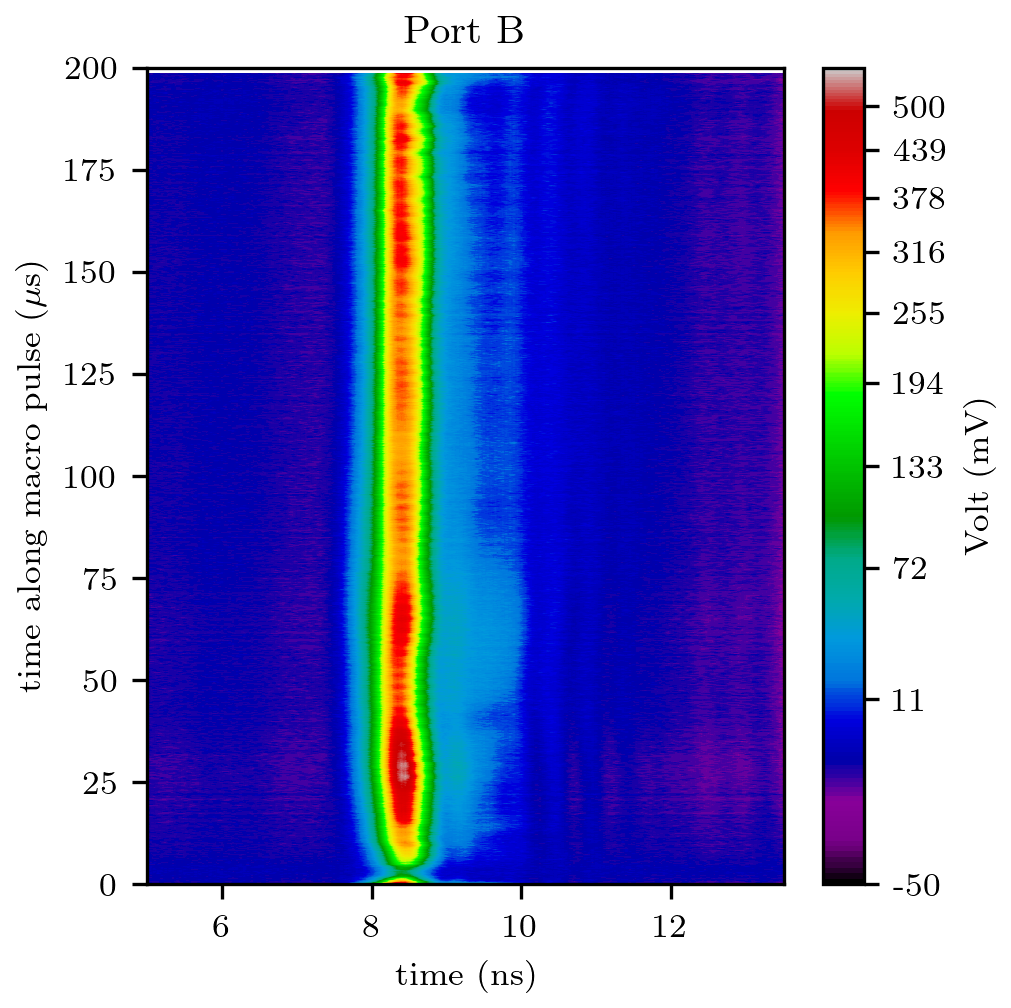}
        \caption{}
        \label{fig-019b}
    \end{subfigure}
    \caption{Waterfall diagram of a full macropulse (about 7200 single bunches) of an \gls{ar10+} beam at \SI{11.4}{\mega\electronvolt\per u} in the experimental cave X2 of \gls{gsi} measured with (\subref{fig-019a}) \gls{rcffc} (position \SI{20.65}{\centi\metre}) and (\subref{fig-019b}) \gls{trcffc} (position \SI{15.85}{\centi\metre}) without biasing.}
    \label{fig-019}
\end{figure}
A full macropulse of about 7200 single bunches is shown in the waterfall diagram in Fig.~\ref{fig-019}, aligned by the RF signal. While the amplitude of the ion peak changes significantly along the macropulse, no strong temporal jitter is observable. No bias voltage was applied to further suppress the \gls{se}. Hence, a light blue tail is visible after the ion peak. This tail is \SI{2}{\nano\second} long for the \gls{rcffc} and \SI{1}{\nano\second} for the \gls{trcffc}. The amplitude of the \gls{se} tail is slightly lower for the \gls{trcffc} and, relative to the signal strength of the ion signal, lower by a factor of \num{2}.

\begin{figure}[!htb]
    \centering
    \begin{subfigure}{0.49\textwidth}
        \includegraphics*[width=.9\columnwidth]{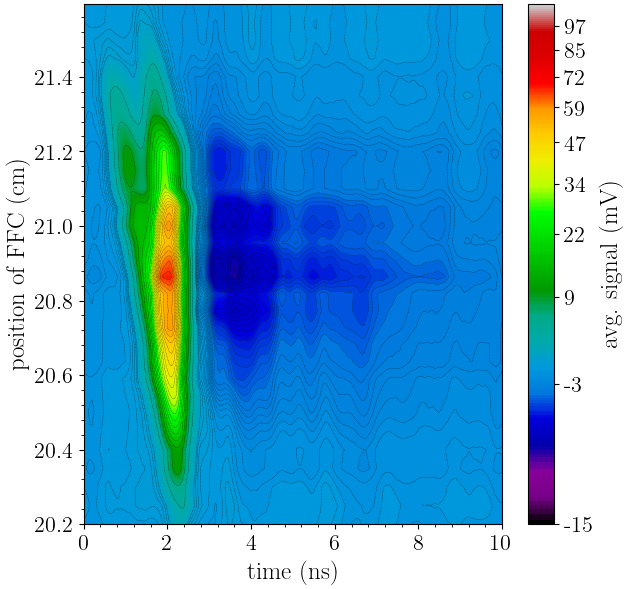}
        \caption{}
        \label{fig-020a}
    \end{subfigure}
    \hfill
    \begin{subfigure}{0.49\textwidth}
        \includegraphics*[width=.9\columnwidth]{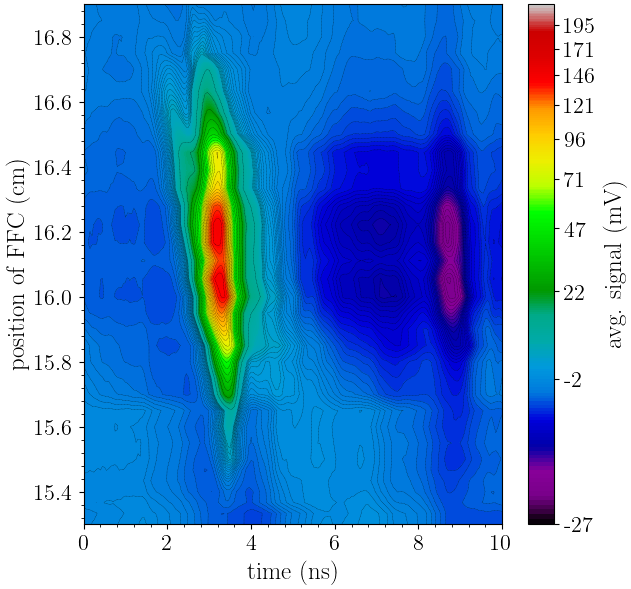}
        \caption{}
        \label{fig-020b}
    \end{subfigure}
    \caption{Average signal of the macro pulse at different insertion positions measured with an \gls{ar10+} beam at \SI{11.4}{\mega\electronvolt\per u} in the experimental cave X2 of \gls{gsi} using (\subref{fig-020a}) \gls{rcffc} and (\subref{fig-020b}) \gls{trcffc} suppression the \gls{se} with a bias voltage of \SI{25}{\volt}.}
    \label{fig-020}
\end{figure}

Due to the pinhole in the heat shield, the FFC selects only a \SI{1.8}{\milli\meter} section in the case of the \gls{trcffc} for typical beams with a transversal extension of about \SI{10}{\milli\metre} in diameter. The most obvious effect is the change in the measured amplitude depending on whether the core of the beam or the outer regions are observed.
The actual shape may alter with respect to the transverse-longitudinal position of the \gls{ffc} due to position dependent effects like dispersion.
In Fig.~\ref{fig-020b}, there is a second peak at \SI{6.9}{\nano\second} at an insertion depth of \SI{16.25}{\centi\metre}, which was not observed at an insertion depth of \SI{15.85}{\centi\metre} as shown in Figures~\ref{fig-018} and \ref{fig-019}. A transversal sweep was performed with both \glspl{ffc} through the beam. A plot of the average bunch shape of the full macropulse against the insertion depth position is shown in Fig.~\ref{fig-020}. A bias of \SI{25}{\volt} has been applied during both measurements. With both \glspl{ffc}, a preceding peak for insertion depths of \SIrange{20.9}{21.6}{\centi\metre} and \SIrange{16.2}{16.8}{\centi\metre} is observed.
The X2 measurement station is connected to the exit of the UNILAC accelerating structure via a series of strong dipoles providing more than a 90 degree bend. Therefore, a large dispersion is expected to be present at the FFC location, although no beamline optics simulations are available. This large dispersion allows the radially coupled \glspl{ffc} to sample the energy axis via its horizontal motion, and therefore, dispersion-assisted longitudinal emittance can be estimated as discussed in~\cite{Singh2022HIAT}.

A horizontal sweep of the FFC was carried out in small steps to measure the bunch lengths. The resolution of this measurement depends on the hole size in the heat shield. Therefore, the finer pinhole of the \gls{rcffc} results in a finer measurement compared to the \gls{trcffc}. In this measurement, the measured $f_\text{gain}$ between both types of \glspl{ffc} can also be evaluated. The maximum intensity of the average bunch shape observed with the \gls{rcffc} is \SI{68.07}{\milli\volt}, while the \gls{trcffc} signal maximum is \SI{145.25}{\milli\volt}.
The measured $f_\text{gain}$ between both \glspl{ffc} is \num{2.13}, which roughly matches a beam width \gls{sigmabtrans} of \SI{2.8}{\milli\metre} according to Eq.~(\ref{eq-2}).

\begin{figure}[!htb]
    \centering
    \begin{subfigure}{0.49\textwidth}
        \includegraphics*[width=.95\columnwidth]{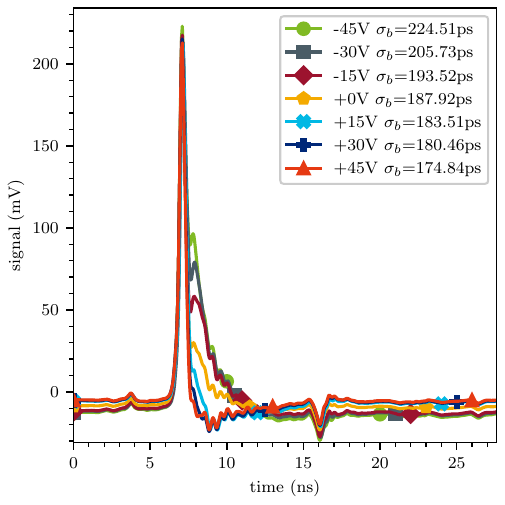}
        \caption{}
        \label{fig-021a}
    \end{subfigure}
    \hfill
    \begin{subfigure}{0.49\textwidth}
        \includegraphics*[width=.95\columnwidth]{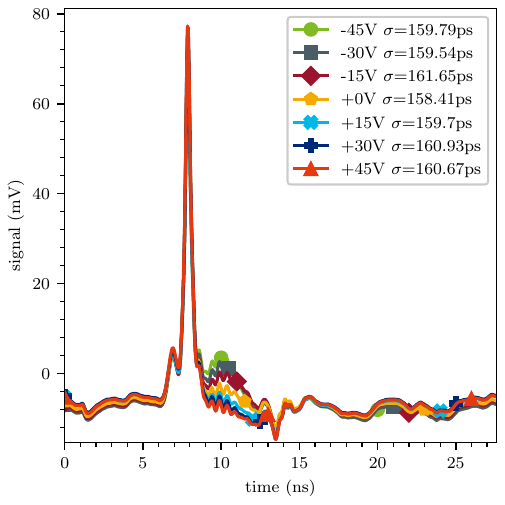}
        \caption{}
        \label{fig-021b}
    \end{subfigure}
    \caption{Average signal of the macro pulse applying different bias voltages measured with an \gls{ar10+} beam at \SI{11.4}{\mega\electronvolt\per u} in the experimental cave X2 of \gls{gsi} using (\subref{fig-021a}) \gls{rcffc} (position \SI{20.65}{\milli\metre}) and (\subref{fig-021b}) \gls{trcffc} (position \SI{16.25}{\milli\metre}). In addition, the bunch length \gls{sigmab} is stated as calculated from the FWHM.}
    \label{fig-021}
\end{figure}

Finally, a bias voltage study was performed for both FFCs. The biasing scheme was applied at an insertion depth of \SI{16.25}{\centi\metre} for the \gls{trcffc}, while it was applied at an insertion depth of \SI{20.65}{\centi\metre} for the \gls{rcffc}, where the second peak is not visible in the scan as shown in Fig.~\ref{fig-020a}. To emphasize the difference in the geometrical \gls{se} suppression of both designs, the bunch shape along the macropulse in Fig.~\ref{fig-021} was averaged for different bias voltages ranging from \SIrange{-45}{45}{\volt}. The bunch width \gls{sigmab} is estimated using the FWHM divided by \num{2.3548}. If the bias voltage is sufficiently high, the bunch length estimation stabilizes at a certain value, even if the bias voltage is increased further. The estimated bunch length changes from \SIrange{224.5}{174.8}{\pico\second} depending on the used bias for the \gls{rcffc}. The temporal separation and the geometrical suppression are not strong enough to suppress the \gls{se} without a bias. A higher bias would be necessary for an accurate estimation. In contrast, the bunch width estimated with the \gls{trcffc} ranges from \SIrange{159.8}{160.9}{\pico\second} at the insertion position of \SI{16.25}{\centi\metre}. Even a high negative bias has no severe impact on the estimated \gls{sigmab}. Hence, the bunch length estimation can be considered to be accurate with an average bunch length over all bias measurements of $\num{160.1}\pm\SI{0.4}{\pico\second}$.

\section{Summary}

This study presented the design, simulation, and experimental validation of Fast Faraday Cups with an emphasis on a newly developed tapered radially coupled \gls{ffc}. The primary objective was to improve the suppression of \glspl{se} and to enhance the signal-to-noise ratio for bunch-by-bunch measurements.

Comparative simulations and RF characterizations of axially and radially coupled FFCs revealed that radial coupling, when combined with a deep, conical drill-hole geometry in the collector, provides superior SE suppression. This geometrical approach allowed effective decoupling of the \gls{se} signals from the primary ion-induced signal, which reduces temporal overlap. The TRCFFC design exhibited a geometrical reduction of \gls{se} down to \SI{2.4}{\percent} of all emitted \gls{se} and can temporally separate \SI{99}{\percent} of the \gls{se} signal from a bunch with a width of $\gls{sigmab}=\SI{400}{\pico\second}$ even without biasing. In general, a bias should be applied and is especially foreseen for high charge state ions.

Additive manufacturing proved highly suitable for fabricating high-precision FFC collectors. Despite slight dimensional deviations and moderate surface roughness, printed copper collectors achieved excellent RF performance surpassing conventionally machined ones. Simulations and S-parameter measurements confirmed that even with the addition of alignment glass pins, the operational bandwidth remained sufficient (up to 7 GHz) for the \gls{trcffc}.

Experimental measurements at the GSI UNILAC further validated the performance. The \gls{trcffc} provided higher signal amplitudes by a factor $\approx 2.13$ compared to the \gls{rcffc} and showed shorter \gls{se} tails. In case of an \gls{ar10+} at \SI{11.4}{\mega\electronvolt\per u}, the measured bunch length remained stable at \SI{160}{\pico\second} for various bias settings, while the RCFFC showed a significant dependency on the bias voltage, which indicates a insufficient \gls{se} suppression under these circumstances. The use case for the \gls{rcffc} are high-bandwidth driven measurements above \SI{3}{\giga\hertz} bandwidth and measurements of slower bunches ($\leq \SI{10}{\percent}~\gls{c0}$), because of the \gls{beta} dependency of the temporal resolution of radially coupled \glspl{ffc}. Additionally, horizontal scans through the beam showed that the radially coupled \glspl{ffc} could detect spatial variations in the bunch structure, which can be caused by e.g. dispersion. The ability to perform dispersion-assisted longitudinal emittance measurements with radially coupled FFCs could be a useful and fast longitudinal emittance measurement technique. Using the FFCs, also further secondary electron emission spectra and yield could be studied.





\vspace{6pt}


\authorcontributions{

    SK and RS contributed to conception, simulations, measurements and writing of the manuscript. SG printed the TRCFFC and performed the roughness measurements. LS, AP and HDeG reviewed the manuscript.}

\funding{This work is supported by the German Federal Ministry of Research, Technology and Space (BMFTR) under contract no. 05P21RORB2. Joint Project 05P2021 - R\&D Accelerator (DIAGNOSE).}

\institutionalreview{
    Not applicable
}

\dataavailability{The data presented in this study are available on request from the corresponding author.}

\acknowledgments{We acknowledge the help from GSI colleague Michael Mueller for help is construction and assembly of the TRCFFCs and GSI mechanical workshop for manufacturing the FFC body and roughness measurements.}

\conflictsofinterest{
    The funders had no role in the design of the study; in the collection, analyses, or interpretation of data; in the writing of the manuscript; or in the decision to publish the results
}



\abbreviations{Abbreviations}{
    The following abbreviations are used in this manuscript:\\

    \printglossary[type=main, title=Glossary]
    \printglossary[type=\acronymtype, title=List of Acronyms]

}

\begin{adjustwidth}{-\extralength}{0cm}

    \reftitle{References}




    \PublishersNote{}
\end{adjustwidth}
\end{document}